# Snowveil: A Framework for Decentralised Preference Discovery

**Grammateia Kotsialou***
King's College London
Grammateia.1.kotsialou@kcl.ac.uk

## Abstract

Aggregating subjective preferences in social choice traditionally assumes a trusted central authority. In contrast, this paper formalises Decentralised Preference Discovery (DPD): the reliable identification of a social choice parameter (e.g. the *canonical outcome* of an aggregation rule applied to the global preference profile) under conditions of partial information, asynchronous interaction, censorship resistance, and no central coordinator. To address DPD, we propose **Snowveil**, a gossip-based framework where agents repeatedly sample random peer rankings and update local beliefs to converge on the canonical outcome. Using a potential function, submartingale theory, and concentration bounds, we prove the system reaches this stable state with tunable high probability, in finite expected time. This single-winner process can then be iterated to construct a set of winning candidates for multi-winner scenarios. Snowveil is agnostic to specific aggregation rules, requiring only that the rule satisfies axioms such as Positive Responsiveness, thus offering a formal basis for a wider class of DPD protocols. Demonstrating Snowveil's modularity, we introduce the *Constrained Hybrid Borda (CHB)*, an aggregation rule designed to balance broad consensus with plurality support. We provide an axiomatic analysis of CHB and present empirical results via extensive simulation, validating Snowveil's $O(n)$ scalability. Overall, this work provides a foundation for how a stable consensus emerges from subjective, expressive, and diverse preference profiles in large-scale decentralised systems.

## 1 Introduction

Decentralised systems facilitate coordination in the absence of a central authority [Wang *et al.*, 2020]. While robust decentralised protocols exist for establishing consensus on replicated state (e.g. objective facts), the problem of aggregating subjective preferences in such environments remains unresolved. This requires designing a mechanism for reaching agreement on social choice outcomes without a trusted coordinator, highlighting a fundamental divergence in the assumptions of established paradigms.

Computational social choice provides axiomatic foundations for preference aggregation [Brandt *et al.*, 2016]; however, standard models conventionally rely on a central authority for ballot collection. Conversely, distributed consensus protocols (e.g. [Nakamoto, 2008; Castro and Liskov, 1999; Yin *et al.*, 2019; Rocket *et al.*, 2019]) achieve agreement in decentralised settings, yet they are designed to replicate objective state by validating transitions against deterministic rules (e.g. digital signatures). This paradigm is fundamentally distinct from preference aggregation. Unlike a distributed ledger transaction, a single preference ranking possesses no intrinsic validity that can be verified in isolation. Instead, its "correctness" is derived solely from the unobservable global preference profile. Consequently, achieving agreement on subjective preferences in this setting necessitates a shift from deterministic verification, typical of classical consensus protocols, to probabilistic estimation.

This shift redefines the protocol's objective function. Standard consensus protocols are designed to minimise latency, such that when there is a disagreement, they force the network to pick a winner as quickly as possible, prioritising *decisiveness* (everyone agreeing) over *accuracy* (picking the correct option). For example, in Nakamoto consensus [Nakamoto, 2008], the longest-chain rule resolves competing branches of the same length by discarding one history to enforce a single path. Similarly, the Snow family of consensus protocols (including Slush, Snowflake, Snowball, and Avalanche in [Rocket *et al.*, 2019]) employs a "snowball effect" mechanism that rapidly tips the network towards the first option to gain a supermajority, thus prioritising early momentum. While the logic presented in both examples is designed to mitigate double-spending, in the context of social choice it would treat a close election as a deadlock to be broken arbitrarily (i.e., the protocol would prioritise decisiveness over aggregate preference accuracy). In turn, this risks the true collective will of an electorate being overridden, since the protocol's rapid conflict resolution would select an outcome before sufficient evidence of support accumulates.

In this paper, we formalise the challenge of aggregating subjective preferences in decentralised systems as *Decentralised Preference Discovery (DPD)*. Unlike consensus on replicated state, which enforces a consistent history of events, DPD requires an electorate to reliably identify a social choice parameter - e.g., the canonical outcome $p^*$ of an aggregation rule $F$ - derived from the global aggregation of private, subjective inputs. The objective is not merely agreement, but the discovery of $p^*$ subject to the inherent constraints of a decentralised setting: partial information, asynchrony, adversarial interference and the absence of a central coordinator. We then bridge the gap between computational social choice and distributed systems by proposing *Snowveil*, a mechanism designed to address DPD. Snowveil is a protocol inspired by the scalable, gossip-based architecture of the Snow family, but replaces their momentum-driven tipping with an axiomatic, preference-based discovery mechanism, i.e. we reformulate the Snow family for the domain of social choice (see Table 1 in supplementary material for comparison). We generalise the consensus payload from objective facts to subjective preferences: instead of verifying binary validity, agents in Snowveil repeatedly sample rankings of their peers, treating these samples not as triggers for rapid tipping, but as noisy local estimates of the emergent global will. By iteratively refining the local state using a locally computable aggregation rule until a confidence threshold is met, the electorate converges to the canonical outcome $p^*$ with high probability.

Our contributions are: $(i)$ we formally define the problem class of DPD; $(ii)$ we introduce Snowveil, a modular gossip-based solution framework for DPD, and establish Positive Responsiveness (PR) as a sufficient condition for liveness; $(iii)$ as part of the protocol design, we introduce the Constrained Hybrid Borda (CHB) aggregation rule, designed to blend broad consensus with plurality support, while also allowing a community to tune the CHB parameters to shift the outcome between the two objectives; $(iv)$ we show that CHB satisfies Fine-Grained Responsiveness (FGR) ensuring it is sensitive enough to act on noisy local samples; $(v)$ modelling Snowveil as a time-homogeneous Markov chain, we use a potential function and submartingale theory to prove that Snowveil converges to a stable state; $(vi)$ we demonstrate, via concentration bounds, that this stable state is identical to the canonical outcome with tunable high probability; $(vii)$ we prove robustness against burying attacks by establishing a linear $\Omega(n)$ lower bound on the coalition size required for strategic manipulation; $(viii)$ we theoretically prove Snowveil exhibits $O(n)$ scalability, driven by a core $n$-independent local probability, and $(ix)$ validate the theoretical findings through extensive simulations.

## 2 Background and Related Work

**Decentralised Consensus and Governance:** Byzantine Fault Tolerant (BFT) protocols, including classical designs (e.g. PBFT [Castro and Liskov, 1999]) and more recent linear-communication variants (e.g. Hotstuff [Yin *et al.*, 2019]) ensure consensus on replicated state but face scalability constraints in large permissionless networks. Alternatively, gossip-based "epidemic" algorithms [Demers *et al.*, 1987] offer robust scalability for numeric aggregation [Kempe *et al.*, 2003a] and distributed polling [Le Merrer *et al.*, 2010], recently extended to asynchronous rank-based statistics [Van Elst *et al.*, 2025]. These approaches focus on statistical estimation - treating variance as error to be filtered - rather than the axiomatic guarantees of social choice, which require fair aggregation of conflicting preferences. The DPD problem is also distinct from decentralised matching markets [Shah *et al.*, 2024], which rely on objective, locally verifiable signals (e.g. acceptance or rejection of proposals). DPD, by comparison, concerns outcomes that cannot be locally verified: correctness depends entirely on the unobservable global preference profile.

Even though gossip protocols are broadly used for transaction and block propagation in blockchain systems (e.g.[Kiffer *et al.*, 2021]), the Snow family [Rocket *et al.*, 2019] is distinctive in leveraging gossip to drive probabilistic consensus as follows: agents repeatedly subsample $k$ peers, creating a positive feedback loop that rapidly tips the network from an unstable equilibrium towards a stable agreement state, with recently formalised consistency and convergence bounds [Amores-Sesar *et al.*, 2024; Buchwald *et al.*, 2026].

In DAO governance, platforms (e.g. Snapshot [Monteiro *et al.*, 2024]) enable expressive voting rules; however, they rely on 'off-chain' aggregation - processing votes on centralised servers to avoid blockchain fees - which reintroduces reliance on a trusted third party. Snowveil eliminates this censorship risk by unifying these distinct paradigms, reformulating the Snow protocols for decentralised, axiomatically-grounded preference discovery.

**Aggregation and Uncertainty in Social Choice:** Computational social choice provides axiomatic foundations for aggregation [Brandt *et al.*, 2016], such as *positive responsiveness* [May, 1952; Horan *et al.*, 2019; Moulin, 1988] and the inevitability of strategic manipulation [Gibbard, 1973; Satterthwaite, 1975]. Beyond deterministic aggregation, the 'probabilistic' view models votes as noisy observations of a ground truth [Mallows, 1957; Doignon *et al.*, 2004; Brandt, 2017]. This perspective is extended to maximum likelihood estimation [Elkind and Shah, 2014], winner determination under uncertainty [Ping and Stoyanovich, 2023], and aggregating partial rankings in peer grading [Caragiannis *et al.*, 2015; Caragiannis *et al.*, 2016; Caragiannis and Micha, 2017]. Structural restrictions on preferences have also simplified the computational burden [Elkind *et al.*, 2022]. These approaches typically assume a central authority to collect and merge the partial or noisy signals. Snowveil eliminates the central aggregator (via gossip) and guarantees convergence through positive responsiveness rather than parametric noise.

Expressive governance paradigms, such as liquid democracy and opinion diffusion, often rely on explicit graphs [Caragiannis and Micha, 2019; Utke and Schmidt-Kraepelin, 2023; Kahng *et al.*, 2021; Halpern *et al.*, 2023; Kotsialou and Riley, 2020; Dhillon *et al.*, 2023; Brill *et al.*, 2022; Revel *et al.*, 2022; Zhang and Grossi, 2022; Colley *et al.*, 2020; Colley *et al.*, 2022; Colley and Grandi, 2022; Kempe *et al.*, 2003b]. Meanwhile, quadratic voting requires pricing mechanisms to capture preference intensity [Lalley and Weyl, 2018]. Snowveil avoids both structural and economic dependencies, focusing instead on the aggregation process itself using graph-free, randomised sampling. Snowveil also diverges from AI-augmented deliberation (e.g. Polis, Generative Social Choice), which is centralised [Revel and Pénigaud, 2025].

## 3 Preliminaries[1] and Problem Formulation

Let $\mathcal{P} = \{p_1, \dots, p_m\}$ be a fixed set of $m$ candidates and $N = \{1, \dots, n\}$ be an electorate of $n$ agents (terms 'agents' and 'voters' are used interchangeably). Each agent $i \in N$ privately holds an initial full strict preference ranking $R_i$ over $\mathcal{P}$. This ranking is defined via a permutation $\pi_i$ of the candidate indices $\{1, \dots, m\}$ such that $R_i = [p_{\pi_i(1)}, p_{\pi_i(2)}, \dots, p_{\pi_i(m)}]$, where $\pi_i(r)$ denotes the index of the candidate ranked in the $r$-th position. The global (full) preference profile is denoted by $\Pi = (R_1, \dots, R_n)$. We assume that $\Pi$ is distributed; that is, no individual agent has full access to $\Pi$, limiting each agent $i$ to a state of partial information restricted to $R_i$.

**Definition 3.1** (Canonical outcome $p^*$)**.** Given a global preference profile $\Pi$ and an aggregation rule $F$, the canonical outcome $p^*$ is the result (e.g. a candidate, a ranking, or a committee) selected by applying the rule $F$ to profile $\Pi$, i.e., $p^* = F(\Pi)$.

**Definition 3.2** (Decentralised Preference Discovery (DPD))**.** We define DPD as the problem of designing a protocol $\mathcal{A}$ such that, for a global preference profile $\Pi$, the electorate reliably identifies a targeted social choice parameter (e.g. the canonical outcome $p^*$) subject to the constraints:

(i) **Partial information:** Each agent $i$ privately holds only ranking $R_i$ and cannot directly access the global preference profile $\Pi$.

(ii) **Asynchrony:** Agents operate without synchronised rounds or a global clock (e.g. communication delays are random).

(iii) **Censorship resistance:** The protocol is robust against adversarial interference.

(iv) **No central coordinator:** There is no central entity collecting ballots or computing the outcome.

The set of constraints in Definition 3.2 is minimal in the sense that relaxing any single one reduces DPD to a trivial or well-solved problem. Specifically, without (i), DPD reduces to local computation; without (ii), it allows for deterministic coordination via synchronous rounds (e.g. flooding); without (iii), it reduces to distributed preference aggregation among cooperative agents, as no mechanisms are needed to defend against interference; without (iv), it reduces to a structurally centralised client-server model. Thus, the DPD problem exists strictly at the intersection of all four constraints, which represents a challenging and general setting.

[1] Proof sketches are presented for main results. All full proofs are provided in the supplementary material.

## 4 The Snowveil Framework

Snowveil provides a constructive solution to DPD by enabling the electorate to reliably identify the canonical outcome $p^*$ through iterative peer-to-peer sampling over an unstructured overlay network. We focus on the case where $F$ is resolute, meaning the core protocol discovers a unique single winner. Snowveil leverages this single-winner mechanism as a sequential building block to construct a full social ranking.

### 4.1 Protocol Overview

Snowveil operates through an $m$-staged convergence process to produce a full ranking outcome over the candidate set $\mathcal{P}$. The only input required from each participant $i \in N$ is the ranking $R_i$. The protocol then constructs the social ranking as follows. Step 1 (Primary discovery): Snowveil instantiates to discover the first-ranked winning candidate, say $w_1$, by engaging the electorate in decentralised sampling until a unique winner is reached via a required quorum. Step $r$ (Sequential elimination): For each subsequent rank $r \in \{2, \dots, m-1\}$, the previous winners are removed from the set of available candidates. The algorithm then re-evaluates this reduced candidate profile using the same underlying individual preferences to decide on the next winner $w_r$. Step $m$: The final remaining candidate is automatically assigned the last position in the ranking, $w_m$. The iterative removal of winners transforms a single-winner discovery engine into a multi-winner ranking framework while maintaining consistent process integrity at every stage.

### 4.2 Dynamic State Model

The protocol's state evolves over discrete time steps. We note that while the system implementation is fully asynchronous (satisfying constraint (ii) of Definition 3.2), we model the evolution using logical time steps $t$. At each time step $t = 0, 1, 2, \dots$, every voter $i \in N$ has a state $s_i(t) \in \mathcal{P} \cup \{\perp\}$, indicating whether they are LOCKED (i.e., they have decided) on a particular candidate within $\mathcal{P}$ or remain UNLOCKED ($\perp$), i.e. remain undecided. The system state is the vector of all voter states, $S_t = (s_1(t), \dots, s_n(t))$. From any state $S_t$, we derive: $(i)$ the set of voters locked on candidate $p_j$: $L_j(t) = \{i \in N \mid s_i(t) = p_j\}$, $(ii)$ the number of locked voters on $p_j$: $N_j(t) = |L_j(t)|$, and $(iii)$ the set of unlocked voters: $U(t) = \{i \in N \mid s_i(t) = \perp\}$.

An iteration of the protocol terminates when the system reaches an *absorbing state*. This occurs when $N_j(t) \geq \lceil Q \cdot n \rceil$ for some candidate $p_j$, where $Q \in (1/2, 1]$ is a predefined global quorum, with $Q > 1/2$ enforcing a unique winner.

### 4.3 The CHB Aggregation Rule

We introduce the **Constrained Hybrid Borda (CHB)** rule, denoted $F_{\text{CHB}}$, to instantiate $F$. The canonical outcome is thus $p^* = F_{\text{CHB}}(\Pi)$. CHB balances broad consensus with plurality support. Given a sample profile $\Pi_k$ of size $k$, let $B(p_j)$ be the Borda score and $t_j$ the first-place support for a candidate $p_j$, where $t_j = |\{R_i \in \Pi_k \mid \pi_i(1) = j\}|$. CHB employs three tunable parameters - $\alpha, \beta, \lambda$ - that function as legitimacy filters. **Popularity Filter** $\alpha \in [0, 1]$**:** a candidate should be $\alpha$-popular ($t_j \geq \lceil \alpha \cdot k \rceil$) to be viable. **Consensus**

**Filter** $\beta \in [0, 1]$**:** a winner's Borda score is at least $\beta$ of the maximum score in the sample. **Hybrid Weight** $\lambda \in [0, 1]$**:** balances consensus ($\lambda = 0$) vs plurality ($\lambda = 1$) in the final score calculation.

The winner is selected via the following sequential selection: (1) **Borda check:** If the candidate with the maximum Borda score is also $\alpha$-popular, they win immediately. (2) **Hybrid selection:** Otherwise, form a set of eligible candidates satisfying both $\alpha$-popularity and $\beta$-consensus. From this eligible set, the winner maximises the hybrid Score $H(p_j)$,

$$H(p_j) = (1 - \lambda) \cdot \underbrace{\left( \frac{B(p_j)}{k \cdot (m - 1)} \right)}_{\text{Normalised Borda Score}} + \lambda \cdot \underbrace{\left( \frac{t_j}{k} \right)}_{\text{Normalised Plurality}} .$$

(3) **Default Case:** If the eligible set is empty, the original Borda winner wins. Ties at any stage break using a deterministic lexicographical rule. This design mitigates extremes, avoiding winners that are either broadly acceptable but lack strong support, or highly polarising.

### 4.4 The Voter Update Process

While the CHB rule selects a winner from a single sample, an unlocked voter needs to verify its stability before committing. To do this, they run the `Voter Update Process` (Algorithm 1), which executes CHB for up to $\gamma$ rounds with two thresholds. The first, $\tau_{\max}$, allows early exit: if any candidate wins $\tau_{\max}$ rounds before $\gamma$ is reached, the voter immediately locks on that candidate, indicating strong local consensus. Otherwise, the process continues for all $\gamma$ rounds. At the end, the second threshold, $\tau_{\min}$, acts as a minimum support filter: the voter locks only if a candidate wins at least $\tau_{\min}$ rounds. Otherwise, they return a `NO-LOCK` outcome, which signals local contention, necessitating abstention until the global state further evolves. To break ties among candidates who have the highest final win-count after all $\gamma$ rounds are complete, the protocol uses recency, assuming that later-round outcomes better reflect the current network state.

### 4.5 Voter Commitment and Protocol Execution

When the `Voter Update Process` returns a `LOCK` decision on a candidate $w$, voter $i$ transitions from the Unlocked to the Locked state. This commitment alters the data shared through peer interactions: if voter $i$ is subsequently sampled by another voter, the locked voter $i$ responds with a temporary ranking $R'_i$, constructed by elevating $w$ to the first position of their private $R_i$ while preserving the order of the remaining candidates. This ensures that $w$ receives maximal Borda and Plurality scores in the sampling peer's aggregation. This process drives a *positive feedback loop*. The Locked state of $i$ persists until the global quorum $Q$ is reached. We note that the underlying $R_i$ remains immutable: once a global winner is reached, all agents revert to (a reduced set of) $R_i$ for the next discovery step. This ensures each iteration round is an independent aggregation based on sincere preferences preventing cross-round strategic effects. We adopt this 'static' preference model also for analytical tractability and clear convergence guarantees. While fully adaptive rankings could accelerate convergence, they introduce dependencies that complicate the analysis. Algorithm 2 formalises the overall execution flow, coordinating the multi-stage convergence to derive the final ranking.

**Algorithm 1 :** `Voter Update Process`

**Require:** Voter $i$, set $\mathcal{P}$, set $N$, parameters $k, \gamma, \tau_{\max}, \tau_{\min}$.
**Initialize:** Set a counter $C[p] \leftarrow 0$ for all $p \in \mathcal{P}$.
**for** round $= 1$ **to** $\gamma$ **do**
  $\Pi_k^{\text{round}} \leftarrow \text{Sample}(N \setminus \{i\}, k)$ ▷ Sample $k$ other voters
  $w_{\text{round}} \leftarrow F_{\text{CHB}}(\Pi_k^{\text{round}})$ ▷ Determine round winner via rule $F_{\text{CHB}}$
  $C[w_{\text{round}}] \leftarrow C[w_{\text{round}}] + 1$
  **if** $C[w_{\text{round}}] \geq \tau_{\max}$ **then**
    **return** `LOCK`($w_{\text{round}}$) ▷ Early exit on strong majority
  **end if**
**end for**
$C_{\max} \leftarrow \max_{p \in \mathcal{P}} C[p]$
**if** $C_{\max} < \tau_{\min}$ **then**
  **return** `NO-LOCK` ▷ Fail if no candidate has minimum support
**else**
  $\mathcal{P}_{\text{winners}} \leftarrow \{p \in \mathcal{P} \mid C[p] = C_{\max}\}$
  **if** $|\mathcal{P}_{\text{winners}}| = 1$ **then**
    **return** `LOCK`(first($\mathcal{P}_{\text{winners}}$))
  **else**
    $w_{\text{final}} \leftarrow \text{TieBreakByRecency}(\mathcal{P}_{\text{winners}})$
    **return** `LOCK`($w_{\text{final}}$)
  **end if**
**end if**

## 5 Axioms for Liveness and Convergence

Snowveil's viability relies on the aggregation rule $F$ satisfying specific functional requirements. To ensure the iterative discovery process is both principled and convergent, we identify a set of minimal sufficient conditions comprising social choice axioms and system constraints. We demonstrate that CHB satisfies these criteria.

### 5.1 Snowveil Compatibility

An aggregation rule $F$ is Snowveil-compatible if it satisfies three fundamental properties: resoluteness (liveness), positive responsiveness (directional correctness), and computability (operational feasibility). Before formalising these, we define the notion of a preference improvement.

**Definition 5.1** ($p$-improvement)**.** A profile $\mathbf{\Pi}'$ is a *$p$-improvement* over $\mathbf{\Pi}$ if it is obtained by a single voter $i$ replacing their ranking $R_i$ with $R'_i$ such that the candidate $p$ is ranked strictly higher in $R'_i$ than in $R_i$, while the relative order of all other candidates remains unchanged.

The compatibility criteria are as follows:

1. RESOLUTENESS. $F$ maps any profile to a single, unique winner. This property (often termed uniqueness or determinism) ensures the update process always progresses towards a well-defined state.

**Algorithm 2** The Snowveil Protocol (High-Level)

**Require:** Sets $N$ (voters) and $\mathcal{P}$ (candidates), global profile $\Pi$, parameters $k, \gamma, \tau_{\max}, \tau_{\min}, Q, \alpha, \beta, \lambda$.
**Ensure:** A complete ordered list, *FinalRanking*.

Initialise *FinalRanking* $\leftarrow$ [ ]
*CurrentCandidates* $\leftarrow \mathcal{P}$
**for** $rank \leftarrow 1$ **to** $m-1$ **do** ▷ Find a winner for this rank
    Reset state: $s_i \leftarrow \perp$ for all $i \in N$.
    **repeat**
        Randomly select unlocked voters to execute Algorithm 1
    **until** a candidate $w$ has $\geq \lceil Q \cdot n \rceil$ locked voters
    Append $w$ to *FinalRanking*.
    Remove $w$ from *CurrentCandidates*.
**end for**
Append the final remaining candidate to *FinalRanking*.
**return** *FinalRanking*.

**Definition 5.2** (Resoluteness)**.** $F$ is resolute if, for any profile $\mathbf{\Pi}$, there exists exactly one $p \in \mathcal{P}$ such that $F(\mathbf{\Pi}) = p$.

2. POSITIVE RESPONSIVENESS. $F$ reacts rationally to increased support: if a candidate gains support, it should never harm their chances. This property underpins directional consistency in the update process and drives the positive feedback loop essential for stochastic convergence.

**Definition 5.3** (Positive Responsiveness, after May (1952))**.** An aggregation rule $F$ is positively responsive if, for any candidate $p \in \mathcal{P}$ and any profile $\mathbf{\Pi}$:

1. **Monotonicity:** If $F(\mathbf{\Pi}) = p$, then for any $p$-improvement $\mathbf{\Pi}'$ over $\mathbf{\Pi}$, it holds that $F(\mathbf{\Pi}') = p$.
2. **Responsiveness:** There exists a profile $\mathbf{\Pi}$ with $F(\mathbf{\Pi}) \neq p$ and a $p$-improvement $\mathbf{\Pi}'$ over $\mathbf{\Pi}$ such that $F(\mathbf{\Pi}') = p$.

3. COMPUTABILITY. An aggregation rule $F$ is computable if there exists an algorithm that returns the unique winner $F(\mathbf{\Pi})$ in finite time. In Snowveil, this serves as a critical liveness constraint: since $F$ is called iteratively in every node's update loop, guaranteed termination is required to prevent 'hanging' nodes that could stall the global discovery.

## 5.2 CHB Compliance

We show that CHB is Snowveil-compatible.

**Proposition 5.4.** *CHB is computable and resolute.*

*Proof sketch.* Computability follows from the polynomial time complexity of scoring. Resoluteness is guaranteed by a strict lexicographical tie-breaking rule, ensuring every profile maps to a unique winner. □

**Proposition 5.5.** *CHB is positively responsive.*

*Proof sketch.* A $p$-improvement strictly increases a candidate's Borda score while non-decreasing their first-place support, strictly improving their hybrid score. This ensures marginal gains break ties or overcome losing margins. □

## 5.3 Fine-Grained Sensitivity

While Positive Responsiveness ensures directional correctness, the efficiency of Snowveil's stochastic discovery depends on the sensitivity of the aggregation rule to marginal signals. Step-function mechanisms often suffer from 'dead zones' where small input changes fail to trigger output changes. Thus we introduce *Fine-Grained Responsiveness* (FGR) to guarantee CHB avoids this pitfall. By reacting to minimal preference changes despite its threshold-based design ($\alpha$,$\beta$), CHB amplifies weak statistical signals, ensuring the system escapes deadlocks.

**Definition 5.6** (Minimal $p$-improvement)**.** A minimal improvement increases a candidate's Borda score by exactly one point without changing its first-place count.

**Definition 5.7** (Fine-Grained Responsiveness)**.** An aggregation rule is FGR if, for any candidate $p$, there exists a profile where a single minimal $p$-improvement makes it the winner.

**Theorem 5.8.** *CHB exhibits Fine-Grained Responsiveness.*

*Proof sketch.* Despite the presence of hard thresholds, the Hybrid Score component of CHB remains strictly increasing with respect to Borda score. We constructively show that a 'tipping point' profile always exists where a single minimal improvement overcomes the losing margin, ensuring the gradient of the feedback loop never flattens to zero. □

# 6 Convergence

We prove that Snowveil converges almost surely. The analysis is structured in three conceptual levels: $(i)$ a single voter's decision terminates, $(ii)$ the system converges on a single winner, and $(iii)$ the full iterative protocol terminates with a complete ranking. The core analysis focuses on $(ii)$, modeling the system as a discrete-time, finite-state Markov chain. We show that, for any aggregation rule satisfying the criteria in Section 5, the protocol converges with probability 1. Crucially, we distinguish this *liveness* guarantee from *accuracy*: we demonstrate in Section 8 that the resulting winner is the canonical outcome with high probability.

## 6.1 The Convergence Argument

The proof of single-winner convergence rests on a potential function argument. We define the potential function $\Phi(S_t)$ that measures the degree of consensus in the system at time $t$,

$$\Phi(S_t) = \sum_{j=1}^{m} (N_j(t))^2,$$

where $N_j(t)$ is the number of voters locked on candidate $p_j$, where $j \in \{1, \ldots, m\}$, at time $t$. Function $\Phi$ is strictly bounded and increases whenever a voter enters a `LOCK` state. The core of the argument relies on the following key lemmas.

**Lemma 6.1** (Positive Decision Probability)**.** *In any non-terminal state, the probability that a randomly chosen unlocked voter transitions to a* `LOCK` *state is strictly greater than zero.*

**Lemma 6.2** (Amplification of Plurality)**.** *If a candidate $p_a$ has more locked voters than a candidate $p_b$, an unlocked voter is strictly more likely to lock on $p_a$ than on $p_b$.*

The latter is a direct result of Positive Responsiveness. Both lemmas imply a positive drift towards consensus.

**Theorem 6.3** (Almost-Sure Single-Winner Convergence)**.** *The Snowveil process for finding a single winner, when instantiated with any rule satisfying the core axioms, is a time-homogeneous Markov chain that almost surely converges in finite time to an absorbing state where a single winner has reached the required quorum $Q$.*

*Proof Sketch.* The sequence of potentials $\{\Phi(S_t)\}$ is a bounded, strict submartingale because the expected potential at the next step is always greater than the current potential (by Lemma 6.1). By the Martingale Convergence Theorem, this process converges to a stable value. Lemma 6.2 creates a positive feedback loop that makes any state of divided consensus transient and unstable. Since the process cannot remain in a fragmented state, it must reach an absorbing state. □

The complete protocol repeats the single-winner convergence $m-1$ times. Since each stage terminates in finite time, the protocol almost surely produces a full ranking.

## 7 Strategic Resilience

While no non-dictatorial rule is strategy-proof [Gibbard, 1973; Satterthwaite, 1975], Snowveil imposes computational barriers by limiting manipulators' global knowledge. We focus on *burying*, the most potent and mathematically tractable attack on the Borda component, to establish a lower bound on strategic robustness.

We quantify the coalition size required to overcome a truthful Borda winner, showing that this size scales linearly with the electorate. Let $p^*$ be the truthful (canonical) winner, possessing an average per-voter Borda score advantage of $\delta > 0$ over a coalition's target candidate $p_c$, defined as

$$\delta = \mathbb{E}_{i \in N_{\text{truth}}}[B(p^*, R_i) - B(p_c, R_i)].$$

To succeed, a manipulative coalition needs to generate an 'anti-signal' within the $k$-samples to overcome this margin.

**Proposition 7.1** (Minimal Coalition Size)**.** *To neutralise the probabilistic advantage of a truthful Borda winner $p^*$ with a per-voter margin of $\delta$, a coordinated coalitional manipulation requires a coalition of size*

$$c \geq n \cdot (\delta/(\delta + m - 1)),$$

*which is $\Omega(n)$.*

*Sketch.* We assume an optimal malicious coalition attempts to bury $p^*$ in favour of $p_c$. We compute the expected Borda score difference in a random $k$-sample, which is a weighted average of the truthful margin $\delta$ and the malicious margin $(m-1)$. Solving for the coalition size $c$ such that this expected difference is non-positive yields the linear bound. □

This linear cost is amplified by Snowveil's design. The $\gamma$-round robustness mechanism requires a coalition to win not just one biased sample, but a supermajority, making an attack exponentially less likely. In the multi-winner setting, the iterative elimination makes optimal long-term planning computationally prohibitive, as it requires reasoning over $m!$ possible elimination paths. This analysis focuses on Borda manipulation; attacks targeting the CHB-specific $\alpha$ and $\beta$ thresholds are an important future research avenue.

## 8 Scalability

We demonstrate Snowveil's scalability by proving the expected convergence time to a single winner scales linearly with the number of voters. The core of this result is showing that the probability of an individual voter locking on the correct winner (canonical outcome) is lower-bounded by a positive constant independent of $n$.

**Lemma 8.1** (Positive $n$-Independent Lock Probability)**.** *For any non-terminal system state, the probability that a randomly chosen unlocked voter transitions to a* `LOCK` *state on the canonical outcome $p^*$ is lower-bounded by a positive constant $c_2 > 0$, which is independent of the number of voters.*

*Sketch.* The proof relies on the Chernoff-Hoeffding concentration bound, modelling the peer selection as sampling with replacement (i.i.d.), a standard approximation valid for $k \ll n$. We show that, for sufficiently large sample size $k$, a random $k$-sample is a high-fidelity proxy for the global preference profile. To do this, we prove the sample mean concentrates sharply around the true margin, $\delta$. This means the probability of a 'misleading sample' that reverses the true winner becomes exponentially small as $k$ grows. Formally, the error probability is bounded by $\exp(-\Omega(k \cdot \delta^2))$. Since this bound depends only on $k$ and the margin $\delta$, not on $n$, each voter has at least a constant probability $c_2 > 0$ of locking on the correct winner, regardless of the network size $n$. □

**Theorem 8.2** (Expected $O(n)$ Convergence Time)**.** *The Snowveil protocol converges on a single winner in an expected number of steps bounded by $O(n)$.*

*Proof Sketch.* The proof follows from Lemma 8.1. The protocol requires at most $n$ successful `LOCK` events to reach a quorum. Since the probability of a single `LOCK` event is lower-bounded by an $n$-independent constant $c_2$, the expected number of steps per lock is at most $\frac{1}{c_2}$, also a constant. Therefore, the total expected time is $n \cdot \frac{1}{c_2} = O(n)$. □

## 9 Experimental Evaluation

**Setup:** We evaluate Snowveil's performance using a discrete-event simulator to capture asynchronous dynamics across two primary metrics: *Convergence Time*, defined as the number of `VoterUpdate` calls required for a candidate to reach the global quorum, and *Decision Accuracy*, the probability that the decentralised protocol terminates on the canonical outcome. We test the protocol against two generative models: *the Impartial Culture* model, representing maximum heterogeneity, and a *Polarised* model, which simulates an electorate split into factions with opposing preferences to test performance under high contention. Unless specified, we use baseline parameters (detailed in the Technical Appendix in the supplementary material). We simulate convergence dynamics up to the LOCK decision. For computational tractability, we abstract ballots from LOCKED voters: their top choice is preserved to maintain the winning momentum, while remaining ranks are replaced with artificial placeholders, enabling extensive runs for our $\mathcal{O}(n)$ scalability claim. This abstraction also serves as a robustness test: by introducing artificial noise in secondary signals, we stress test the feedback loop driving convergence. As Snowveil performs well even under these conditions, our results represent a conservative estimate on the protocol's true performance.

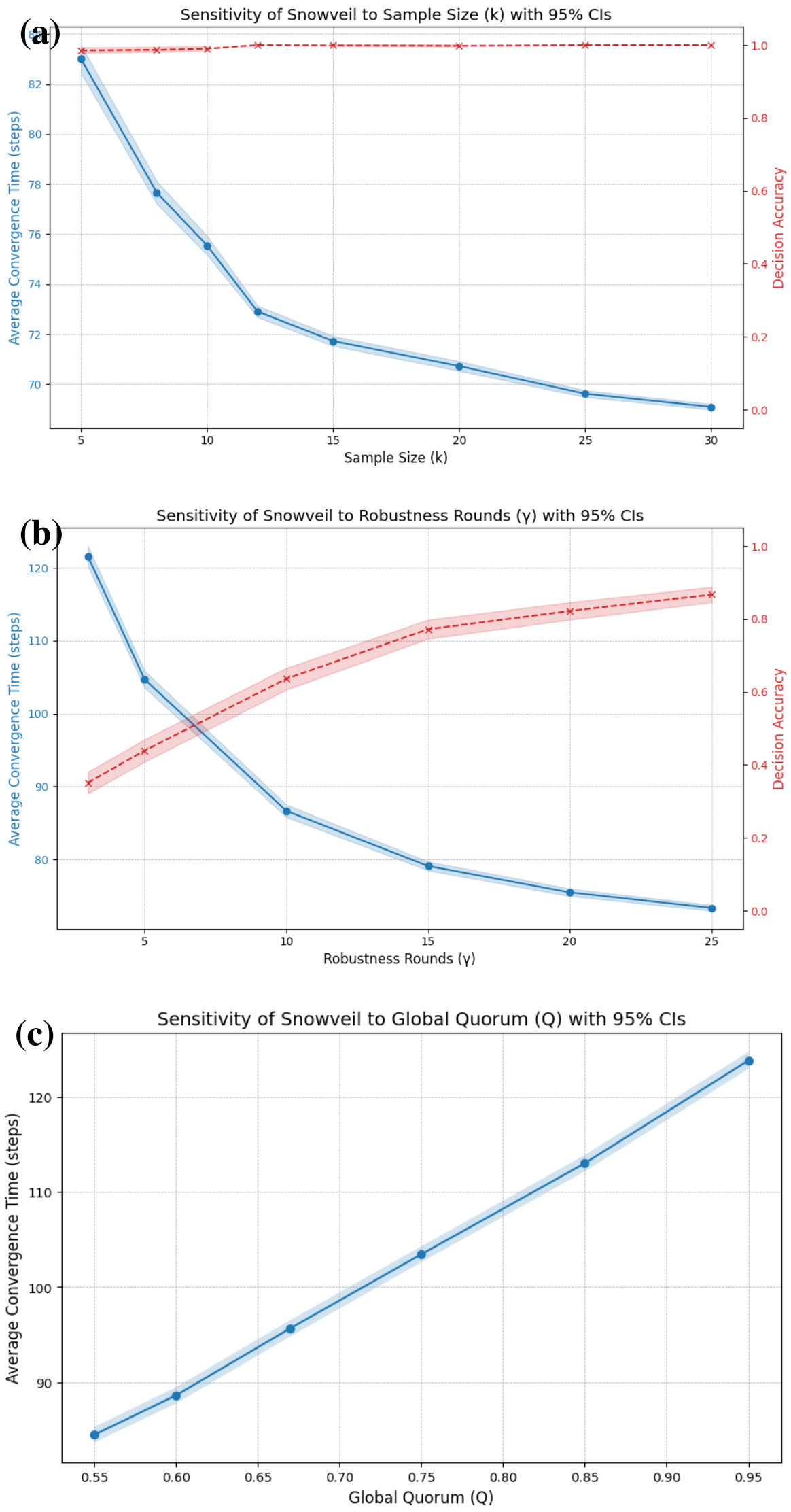

Figure 1: Sensitivity for key parameters: (a) Sample size $k$, (b) robustness rounds $\gamma$, (c) global quorum $Q$. In (a) and (b) plots, the blue line (left axis) denotes average convergence time, while the red dotted line (right axis) represents decision accuracy (fidelity to the canonical winner). Accuracy approaches 1.0 as $k$ and $\gamma$ increase, confirming Snowveil reconstructs the global preference profile.

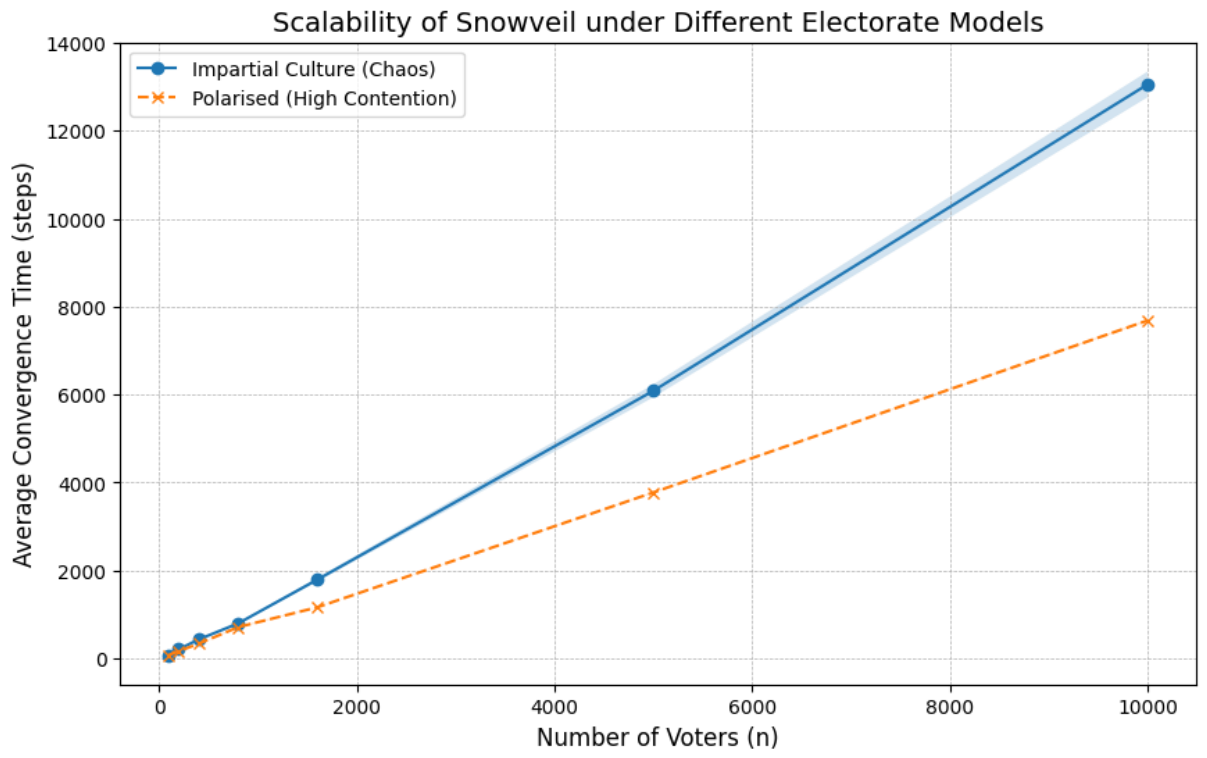

Figure 2: Scalability vs. number of voters $n$ for Impartial Culture and Polarised electorates (95% CI).

**Results:** Figure 1(b) reveals a 'cautious voter paradox': increasing the number of local robustness rounds $\gamma$ significantly decreases global convergence time. Higher-quality local decisions, while individually more costly, reduce network contention and accelerate consensus while also ensure that Decision Accuracy approaches 1.0, as the protocol filters out sampling noise. A similar trend is observed for the sample size $k$ (Figure 1 (a)), where performance gains and fidelity both plateau for $k > 15$, suggesting that larger samples provide a sufficient statistical proxy for the global preference profile. Finally, Figure 1(c) shows that the convergence time grows super-linearly as the quorum $Q$ approaches unanimity. Figure 2 provides empirical validation for Theorem 8.2. Under both the chaotic IC and the high-contention Polarised models, the convergence time scales linearly with $n$, confirming the protocol's $O(n)$ complexity. Interestingly, convergence is faster in the polarised setting, suggesting the protocol's feedback loop effectively amplifies pre-existing social signals to break deadlocks.

## 10 Conclusion and Discussion

We introduce Snowveil, a principled framework for Decentralised Preference Discovery (DPD), demonstrating that stable collective consensus can emerge from subjective preferences in a large-scale, gossip-based system. By integrating stochastic feedback dynamics of iterative sampling with an axiomatic analysis of aggregation rules, Snowveil provides a formal method for decentralised preference aggregation.

We explicitly separate the functional requirements for protocol liveness from the normative properties of the outcome. We identify Positive Responsiveness as a fundamental prerequisite for convergence, as it ensures that the protocol amplifies marginal statistical advantages into network-wide decisions. Fairness in Snowveil is thus tunable via the CHB parameters $\alpha$, $\beta$ and $\lambda$, allowing communities to encode their own social contract, without altering the underlying convergence engine.

Unlike Nakamoto consensus [Nakamoto, 2008], which optimises for agreement on an objective history (deciding which block to append), Snowveil is designed for the multi-alternative domain of subjective preferences. It discovers the canonical outcome, determined by the global preference profile, using only distributed, local views, achieving high probability convergence without a central coordinator. Thus, Snowveil is best viewed as a decentralised governance primitive rather than a ledger protocol, designed to complement rather than compete with existing consensus protocols.

While this work prioritises the introduction of a 'bridge' model between social choice and distributed systems, it opens several research avenues including: addressing challenges posed by dynamic preferences, dynamic participation, privacy preservation, established fairness axioms and weighted ballots.

| Aspect | Snow family protocols (Slush, Snowflake, Snowball) | Snowveil framework |
|---|---|---|
| *I. Objective & Outcome* | | |
| What is proven? | Probabilistic agreement on a binary/$k$-ary state; all correct nodes converge to one "color" (decision). | Staged convergence: almost-sure convergence to a top winner; iteration yields a complete ranking over $m$ alternatives. |
| Outcome space | Single choice from a small, predefined set. | Full permutation (via elimination), with *ranking integrity* across rounds. |
| Termination model | Typically long-running consensus over many decisions. | Per-winner terminating subroutine; outer loop repeats $m-1$ times to build the ranking. |
| *II. Proof Architecture & Methodology* | | |
| Overall approach | Analyses often emphasize local sampling bias and metastability intuition. | Analytical potential-function method: model as a finite, time-homogeneous Markov chain; prove strict submartingale drift. |
| Mathematical model | Discrete-time stochastic process; comparisons to biased random walks are common. | Explicit Markov chain on lock-states; convergence via potential $\Phi(S_t) = \sum_j N_j(t)^2$ and submartingale arguments. |
| Core tool | Binomial/sub-sampling probabilities to show local asymmetry. | Potential functions & martingale theory; identify *Positive Responsiveness (PR)* as a sufficient condition for liveness. |
| *III. Drift Mechanism* | | |
| How drift is shown | If one option has plurality, a node more likely samples it; local adoption probabilities amplify the majority. | Global potential has positive drift (strict submartingale); any fragmented state is transient; absorption at quorum. |
| Level of analysis | Primarily micro-level (a node's sample and update). | Macro/meso-level (system state evolution); micro only needs nonzero progress probability. |
| Role of exact probabilities | Central (explicit probability comparisons). | Abstracted: only need a positive lower bound on lock probability; not the exact values. |
| *IV. Update Rule & State Complexity* | | |
| Update rule | Simple thresholding (e.g., adopt if $\alpha$-majority in sample). | Constrained Hybrid Borda (CHB): combines normalized Borda and plurality with tunable $(\alpha, \beta, \lambda)$. |
| State space | Low (node colors: e.g., $2^n$ for binary). | Higher: vector of voter lock-states over $m$ projects, i.e., $(m+1)^n$. |
| Generality of convergence proof | Typically coupled to the specific update rule/thresholds. | *Pluggable*: any rule satisfying PR + resoluteness + computability inherits the convergence guarantee. |

Table 1: Technical comparison of convergence proofs: Snow-family protocols and Snowveil. Snow protocols pioneered scalable, metastable consensus for objective states; Snowveil builds on these foundations to address subjective preference aggregation, introducing an axiomatic layer (PR-based liveness), a global potential-function argument, and a staged multiple winner process with ranking integrity.

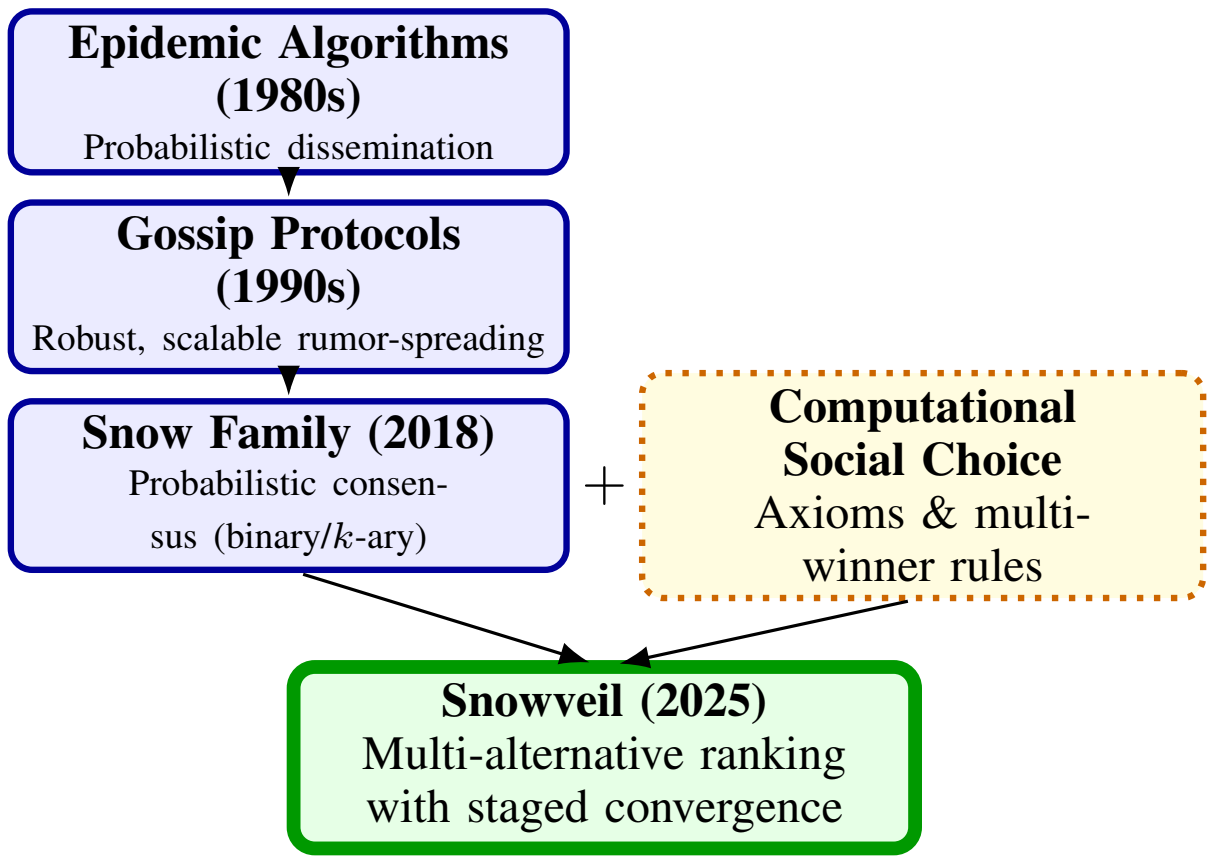

Figure 3: Snowveil as a synthesis of Distributed Systems and Computational Social Choice

## A Proofs of Section 5: Axioms for Liveness and Convergence

**Proposition 5.4.** The Constrained Hybrid Borda (CHB) aggregation rule is computable and resolute.

*Proof.* We prove each property of the rule. We assume an input profile $\Pi_k$ consisting of $k$ strict preference rankings over a set of $m$ candidates $\mathcal{P} = \{p_1, p_2, \ldots, p_m\}$.

**Computability.** To prove CHB is computable, we must show that its outcome can be determined in a finite number of computational steps:

- *Borda Score Calculation:* For each of the $k$ ballots, assigning scores from $m-1$ down to 0 takes $O(m)$ time. Summing these scores for all $m$ candidates across all $k$ ballots takes $O(k \cdot m)$ time.
- *First-Place Support (Plurality) Calculation:* This requires iterating through the top choice of each of the $k$ ranking ballots, which takes $O(k)$ time.
- *Threshold Checks:* Identifying the maximum Borda score ($B^{\max}$), checking the $\alpha$-popularity of the Borda winner, constructing the set $\mathcal{P}_{\text{eligible}}$, and checking if this set is empty are operations that involve a finite number of comparisons and arithmetic operations over the $m$ candidates. These steps are completed in $O(m)$ time.
- *Hybrid Score Calculation:* When required, calculating the Hybrid Score for each of the at most $m$ candidates in $\mathcal{P}_{\text{eligible}}$ involves a fixed number of arithmetic operations per candidate. This also takes $O(m)$ time.
- *Winner Selection:* Finding the maximum score in a list of $m$ items takes $O(m)$ time.

Since every step of the CHB algorithm consists of a finite sequence of arithmetic operations and comparisons on finite sets, the total time complexity is polynomial in the number of agents (which, for $\Pi_k$, number of agents equals to $k$ since $\Pi_k$ is only a $k$-sample size of the full $n$-size global profile $\Pi$) and the number of candidates ($m$). Therefore, the rule is computable.

**Resoluteness.** An aggregation rule is resolute if, for any profile $\mathbf{\Pi}$, it maps to exactly one candidate $p \in \mathcal{P}$. This requires the rule to be both well-defined (deterministic) and to always return a singleton set (uniqueness).

- **Determinism:** An algorithm is deterministic if, for a given input, it always produces the same output. The CHB rule is constructed from deterministic operations: $(i)$ the calculation of the (arithmetic sums) Borda scores and first-place counts, $(ii)$ the comparison of scores against the fixed thresholds ($\alpha$, $\beta$), $(iii)$ the calculation of the Hybrid Score, $(iv)$ and the selection of a maximum value given a set of scores. Since each component of the CHB rule is deterministic, the sequence of operations for a given preference profile $\Pi_k$ is fixed, it follows the same execution path and produce the same numerical scores. Assuming a deterministic tie-breaking rule (see below), the final outcome is also guaranteed to be the same. Therefore, the rule is deterministic.
- **Uniqueness via Tie-Breaking:** To guarantee a unique outcome, we show that a single winner is produced in all possible scenarios, which requires a well-defined tie-breaking rule. Lets assume a standard, deterministic tie-breaking rule, the lexicographic one: if two or more candidates are tied at any decision point, the winner is the candidate with the lowest index (i.e., $p_i$ wins over $p_j$ if $i < j$). Then there are three possible paths to selecting a winner in the CHB algorithm:
  - Case 1: The Borda winner is $\alpha$-popular. The algorithm first identifies the candidate(s) with the maximum Borda score, $B^{\max}$. If a single candidate $p_j$ uniquely has this score, then $p_j$ is selected as $p_{B^{\max}}$. If multiple candidates tie for the highest Borda score, our lexicographical tie-breaking rule is invoked to select a single, unique candidate from the tied set, which we designate as $p_{B^{\max}}$. The algorithm then checks if this unique $p_{B^{\max}}$ is $\alpha$-popular. If yes, $p_{B^{\max}}$ is the declared winner.
  - Case 2: The fallback Borda winner is chosen. This occurs if the (uniquely identified) Borda winner $p_{B^{\max}}$ is not $\alpha$-popular and the set $\mathcal{P}_{\text{eligible}}$ is empty. In this case, the rule defaults to declaring $p_{B^{\max}}$ as the winner. Since $p_{B^{\max}}$ was uniquely determined in the first step (using tie-breaking if necessary), the outcome is unique.
  - Case 3: A Hybrid Score winner is chosen. This case occurs when $\mathcal{P}_{\text{eligible}}$ is non-empty. The algorithm calculates the Hybrid Score $H(p_j)$ for all candidates in $\mathcal{P}_{\text{eligible}}$. If a single candidate has a unique maximum Hybrid Score, it is declared the winner. If multiple candidates in $\mathcal{P}_{\text{eligible}}$ tie for the highest Hybrid Score, our lexicographical tie-breaking rule selects a winner from this tied set. In this case, the outcome is again unique.

  Since all possible execution paths of the algorithm lead to a single unique winner, either by unique merit or by the application of a deterministic tie-breaking rule, the CHB rule is resolute.

This completes the proof. □

**Proposition 5.5.** The Constrained Hybrid Borda (CHB) aggregation rule is *positive responsive*.

*Proof.* **Part $(a)$: Proof of Monotonicity**

Let $\Pi_k$ be an initial profile of $k$ preference rankings (recall that CHB takes as input the rankings from a $k$ in size sample). Let the CHB winner for this profile be $p^{\text{CHB}}$. Let $\Pi'_k$ be a new profile where a single voter $u$ has modified their ranking by moving $p^{\text{CHB}}$ to a higher position, without changing the relative order of any other candidates, $\{p_j\}_{p_j \neq p^{\text{CHB}}}$. Let $B(p_j)$, $t_j$, and $H(p_j)$ denote the Borda score, first-place support, and hybrid score for any candidate $p_j$ under profile $\Pi_k$. Let $B'(p_j)$, $t'_j$, and $H'(p_j)$ be the corresponding scores under profile $\Pi'_k$. Then we analyse the effect of the preference change (by $u$) on these scores. When voter $u$ raises $p^{\text{CHB}}$'s rank, we have:

- the Borda score of $p^{\text{CHB}}$ strictly increases ($B'(p^{\text{CHB}}) > B(p^{\text{CHB}})$),
- the first-place support for $p^{\text{CHB}}$ either increases or stays the same ($t'^{\text{CHB}} \geq t^{\text{CHB}}$):
  - The first-place count remains the same ($t'^{\text{CHB}} = t^{\text{CHB}}$), which happens in two sub-cases:
    1. Voter $u$ already had $p^{\text{CHB}}$ as their first choice. They cannot move it any higher, so their ranking does not change, and the count $t^{\text{CHB}}$ is unaffected.
    2. Voter $u$ did not have $p^{\text{CHB}}$ as their first choice, and their "boost" moves $p^{\text{CHB}}$ up, but not to the first position, e.g. from $[A, B, p^{\text{CHB}}, D]$ to $[A, p^{\text{CHB}}, B, D]$. Their first choice remains A, so their contribution to the first-place count for $p^{\text{CHB}}$ remains zero. The total count $t^{\text{CHB}}$ is unaffected.
  - The First-Place Vote Count Increases ($t'^{\text{CHB}} > t^{\text{CHB}}$):
    1. Voter $u$ did not have $p^{\text{CHB}}$ as their first choice initially (e.g., their ranking was $[A, B, p^{\text{CHB}}, D]$).
    2. Their "boost" involves moving $p^{\text{CHB}}$ to the #1 spot. Their new ranking becomes $[p^{\text{CHB}}, A, B, D]$. In this scenario, the total first-place vote count for $p^{\text{CHB}}$ increases by exactly one, because voter $v$ is now counted in that tally.
- for any other candidate $p_j \neq p^{\text{CHB}}$, its Borda score and first-place support can only decrease or stay the same: $B'(p_j) \leq B(p_j)$ and $t'_j \leq t_j$:
  - First-place support does not increase ($t'_j \leq t_j$): We look at the single ballot that changes.
    1. Voter $u$'s top choice was not $p_j$ to begin with: Their top choice was some other candidate, say $p_A$. When they move $p^{\text{CHB}}$ up, their top choice either remains $p_A$ (if $p^{\text{CHB}}$ is not moved to #1) or it becomes $p^{\text{CHB}}$ (if $p^{\text{CHB}}$ is moved to #1). In neither scenario does $p_j$ suddenly become their #1 choice. So, the count $t_j$ does not increase.

2. Voter $u$'s top choice was $p_j$ to begin with Voter $v$ moves $p^{\text{CHB}}$ up. If they move $p^{\text{CHB}}$ into the #1 spot, their new top choice is $p^{\text{CHB}}$, not $p_j$. In this case, the total count $t_j$ decreases by one.

– Borda score does not increase ($B'(p_j) \leq B(p_j)$): When voter $u$ moves $p^{\text{CHB}}$ up in their ranking, we check what happens to the rank of another candidate $p_j$.

1. $p_j$ was ranked lower than $p^{\text{CHB}}$ to begin with: When $p^{\text{CHB}}$ moves up, it moves away from $p_j$. The rank of $p_j$ does not change relative to the candidates around it. Its rank stays the same, so its Borda score is unchanged.
2. $p_j$ was ranked higher than $p^{\text{CHB}}$ to begin with: When $p^{\text{CHB}}$ is moved up, it might jump over $p_j$. If it does, $p_j$ gets 'pushed down' one spot to a worse rank, implying a lower Borda score. If $p^{\text{CHB}}$ moves up but does not reach $p_j$, then $p_j$'s rank is unaffected.

By the definition of the Hybrid Score, it follows that $H'(p^{\text{CHB}}) > H(p^{\text{CHB}})$, and $H'(p_j) \leq H(p_j)$ for any $p_j \neq p^{\text{CHB}}$].

Let $p^{B^{\max}}$ and $p'^{B^{\max}}$ be the Borda winners in $\Pi_k$ and $\Pi'_k$, respectively. From the score changes, it follows that the maximum Borda score in the new profile cannot be greater than the old one unless $p^{\text{CHB}}$ becomes the new Borda winner, that is, $B'^{\max} \leq B^{\max}$ if $p'^{B^{\max}} \neq p^{\text{CHB}}$, and $B'^{\max} = B'(p^{\text{CHB}}) > B(p^{\text{CHB}})$ if $p'^{B^{\max}} = p^{\text{CHB}}$.

We now proceed by cases on how $p^{\text{CHB}}$ wins under the initial profile $\Pi_k$:

**Case 1: $p^{\text{CHB}}$ won as an $\alpha$-popular Borda winner.** In this case, $p^{\text{CHB}} = p^{B^{\max}}$ and $t^{\text{CHB}} \geq \lceil \alpha k \rceil$. In profile $\Pi'_k$, since $B'(p^{\text{CHB}}) > B(p^{\text{CHB}})$ and $B'(p_j) \leq B(p_j)$ for all $p_j \neq p^{\text{CHB}}$, we have that $p^{\text{CHB}}$ should remain the Borda winner. Thus, $p'^{B^{\max}} = p^{\text{CHB}}$.

Since $t'^{\text{CHB}} \geq t^{\text{CHB}}$, the condition $t'^{\text{CHB}} \geq \lceil \alpha k \rceil$ holds. Thus, in $\Pi'_k$, $p^{\text{CHB}}$ is the Borda winner and is $\alpha$-popular, so it wins again. Monotonicity holds.

**Case 2: $p^{\text{CHB}}$ wins by default as the Borda winner because $\mathcal{P}_{\text{eligible}}$ is empty.** We have that $p^{\text{CHB}} = p^{B^{\max}}$, however $t^{\text{CHB}} < \lceil \alpha k \rceil$, and for all candidates $p_j$, it either holds that $t_j < \lceil \alpha k \rceil$ or that $B(p_j) < \beta \cdot B^{\max}$.

In $\Pi'_k$, $p^{\text{CHB}}$ remains the Borda winner, $p'^{B^{\max}} = p^{\text{CHB}}$. If $p^{\text{CHB}}$ becomes $\alpha$-popular (i.e., $t'^{\text{CHB}} \geq \lceil \alpha k \rceil$), it wins under Case 1.

If $p^{\text{CHB}}$ remains not $\alpha$-popular, we must check the new eligibility set, $\mathcal{P}'_{\text{eligible}}$: A candidate $p_j$ is in $\mathcal{P}'_{\text{eligible}}$ if $t'_j \geq \lceil \alpha k \rceil$ and $B'(p_j) \geq \beta \cdot B'^{\max}$.

For any candidate $p_j \neq p^{\text{CHB}}$, $t'_j \leq t_j$ and $B'(p_j) \leq B(p_j)$. Since $p^{\text{CHB}} = p'^{B^{\max}}$, $B'^{\max} > B^{\max}$. Thus, the conditions to enter $\mathcal{P}'_{\text{eligible}}$ are stricter for any $p_j \neq p^{\text{CHB}}$. No other candidate can become eligible for $\mathcal{P}'_{\text{eligible}}$. Since we assume $p^{\text{CHB}}$ remains not $\alpha$-popular, $\mathcal{P}'_{\text{eligible}}$ is empty. Thus the rule defaults to the Borda winner $p^{\text{CHB}}$. Monotonicity holds.

**Case 3: $p^{\text{CHB}}$ won with the highest Hybrid Score from a non-empty $\mathcal{P}_{\text{eligible}}$.** This means $p^{\text{CHB}} \in \mathcal{P}_{\text{eligible}}$, thus $t^{\text{CHB}} \geq \lceil \alpha k \rceil$ and $B(p^{\text{CHB}}) \geq \beta \cdot B^{\max}$. And for any other candidate $p_j \in \mathcal{P}_{\text{eligible}}$, $H(p^{\text{CHB}}) \geq H(p_j)$.

First, we show that $p^{\text{CHB}}$ remains eligible in $\Pi'_k$.

- The popularity condition holds: $t'^{\text{CHB}} \geq t^{\text{CHB}} \geq \lceil \alpha k \rceil$.
- The Borda condition: We must show $B'(p^{\text{CHB}}) \geq \beta \cdot B'^{\max}$.
  - If $p^{\text{CHB}}$ becomes the new Borda winner, $p'^{B^{\max}} = p^{\text{CHB}}$, the condition holds as $\beta \leq 1$.
  - If the Borda winner is some $p_w \neq p^{\text{CHB}}$, then $B'^{\max} = B'(p_w) \leq B(p_w) \leq B^{\max}$. We have $B'(p^{\text{CHB}}) > B(p^{\text{CHB}}) \geq \beta \cdot B^{\max} \geq \beta \cdot B'^{\max}$. The condition holds.

Thus, $p^{\text{CHB}}$ is in $\mathcal{P}'_{\text{eligible}}$.

Next, we show taht $p^{\text{CHB}}$ still wins. The set of competitors for $p^{\text{CHB}}$ in $\Pi'_k$, $\mathcal{P}'_{\text{eligible}} \setminus \{p^{\text{CHB}}\}$, is a subset of its competitors in $\Pi_k$, because for any $p_j \neq p^{\text{CHB}}$, the conditions for eligibility do not become easier. Let $p_j$ be any other candidate in $\mathcal{P}'_{\text{eligible}}$. We have the following chain of inequalities:

$$H'(p^{\text{CHB}}) > H(p^{\text{CHB}}) \geq H(p_j) \geq H'(p_j).$$

The first inequality holds because $p^{\text{CHB}}$'s scores improved. The second is the condition for $p^{\text{CHB}}$'s victory in $\Pi_k$. The third holds because $p_j$'s scores did not improve. This shows that $H'(p^{\text{CHB}}) > H'(p_j)$ for any other eligible candidate $p_j$.

Note that $p^{\text{CHB}}$'s win is supported by a strict inequality (score increase), thus $p^{\text{CHB}}$ is the unique winner regardless of the deterministic tie-breaking rule. Therefore, $p^{\text{CHB}}$ wins the Hybrid Score comparison again, and monotonicity holds.

Since $p^{\text{CHB}}$ wins in the new profile $\Pi'_k$ in all possible cases, the CHB rule is monotonic. This proves the part $(a)$ of the proof. We proceed to prove the stronger notion of responsiveness.

**Part $(b)$: Proof of Responsiveness** (of Proposition 5.5 in page 11)

We note that if the Positive Responsiveness property only guaranteed that a boost could break a perfect tie, the protocol would be brittle. In the far more common scenario of a narrow loss, there would be no guarantee that a small improvement could flip the outcome. The system could get stuck in a state where one candidate always has a tiny advantage, and the protocol would have no mechanism to escape that 'bad equilibrium' situation. Therefore, a 'tie-breaking only' version of positive responsiveness would be too weak. The stronger 'loser-to-winner' property is essential because it guarantees the protocol can make progress from the far more common states of narrow contention, which is the key to ensuring the liveness and robustness of the entire convergence process.

For this reason, we show that for any candidate $p_j \in \mathcal{P}$, there exists at least one profile $\Pi_k$ such that $F_{CHB}(\Pi_k) \neq p_j$, and a corresponding *$p_j$-improvement* $\Pi'_k$ over $\Pi_k$, for which $F_{CHB}(\Pi'_k) = p_j$. We prove this by cases, each demonstrating how such profiles, $\Pi_k$ and $\Pi'_k$, can be constructed for an arbitrary candidate $p_j$.

Let $p_q$ be another distinct candidate:

**Case 1: Making $p_j$ win by becoming the $\alpha$-popular Borda Winner.** The goal is to construct a profile $\Pi_k$ where $p_j$ is the Borda winner but is not $\alpha$-popular, and then show that a single $p_j$*-improvement* makes $p_j$ win:

- Profile $\Pi_k$: Within a sample of $k$ voters, arrange the ballots such that $p_j$ has the highest Borda score, $B(p_j) = B^{\max}$. This can be achieved by giving $p_j$ many second-place ranks. Set the number of first-place votes for $p_j$ to be exactly one less than the popularity threshold: $t_j = \lceil \alpha \cdot k \rceil - 1$. Arrange the ballots such that another candidate, $p_q$, is $\alpha$-popular ($t_q \geq \lceil \alpha \cdot k \rceil$) and has a high enough Borda score to be in the set $\mathcal{P}_{\text{eligible}}$ and win based on its Hybrid Score. Thus, we have a profile where $F_{CHB}(\Pi_k) = p_q \neq p_j$.
- Profile $\Pi'_k$: Create $\Pi'_k$ from $\Pi_k$ with a single $p_j$-*improvement*: have one voter who previously ranked another candidate first change their ballot to rank $p_j$ first.
- Analyse outcome in $\Pi'_k$: In $\Pi'_k$, $p_j$'s first-place vote count is now $t'_j = t_j + 1 = \lceil \alpha k \rceil$, thus it is now $\alpha$-popular. Its Borda score $B'(p_j)$ also increased, thus it remains the Borda winner, $B'(p_j) = B'^{\max}$. According to CHB, if the Borda winner is $\alpha$-popular, it wins immediately. Therefore, $F_{CHB}(\Pi'_k) = p_j$. The outcome has flipped from $p_q$ to $p_j$.

**Case 2: Making $p_j$ win by winning the Hybrid Score competition.** The goal is to construct a profile $\Pi_k$ where $p_j$ is an eligible candidate but loses narrowly to another eligible candidate, and then show that a marginal improvement makes $p_j$ win:

- Profile $\Pi_k$: We arrange the $k$ ballots to create a 'close race' scenario such that: both $p_j$ and $p_q$ satisfy the conditions for the set $\mathcal{P}_{\text{eligible}}$ (i.e., both are $\alpha$-popular and meet the $\beta$-consensus threshold), and the Borda scores and plurality counts are arranged such that the Hybrid Score of $p_q$ is marginally greater than that of $p_j$: $H(p_q) > H(p_j)$, where the difference $\epsilon = H(p_q) - H(p_j)$ is a small positive value. In this case, $F_{CHB}(\Pi_k) = p_q \neq p_j$.
- Profile $\Pi'_k$: Create $\Pi'_k$ from $\Pi_k$ with a single $p_j$-*improvement*: have one voter modify their ballot to raise the rank of $p_j$. Any action that improves $p_j$'s rank increases the Borda score $B(p_j)$.
- Analyse the outcome in $\Pi'_k$: This $p_j$*-improvement* results in a positive increase to $p_j$'s Borda score and, consequently, a strictly positive increase to its Hybrid Score, $H(p_j)$. Let this increase be $\Delta H > 0$. Meanwhile, note that the score for the other candidate, $p_q$, cannot increase as a result of a $p_j$*-improvement*, so $H'(p_q) \leq H(p_q)$. In our construction of $\Pi_k$, we create a 'close race' with a 'sufficiently small' losing margin, $\epsilon = H(p_q) - H(p_j)$. This allows us to design the profile $\Pi_k$ such that the margin $\epsilon$ is smaller than the positive gain $\Delta H$ that results from our chosen $p_j$*-improvement*. Therefore, we have $H'(p_j) = H(p_j) + \Delta H > H(p_j) + (H(p_q) - H(p_j)) = H(p_q) \geq H'(p_q)$. With its new score, $p_j$ has the highest Hybrid Score in $\mathcal{P}_{\text{eligible}}$, and thus $F_{CHB}(\Pi'_k) = p_j$. The outcome has flipped.

Since we have shown that, for an arbitrary candidate $p_j$, a profile can be constructed where $p_j$ is not the winner, but a single $p_j$*-improvement* makes $p_j$ the winner, the CHB rule satisfies the responsiveness condition. This completes the part $(b)$ of the proof, thus it completes the proof of Proposition 5.5, stating that CHB is positively responsive. □

**Theorem 5.8.** The Constrained Hybrid Borda (CHB) aggregation rule exhibits fine-grained responsiveness.

*Proof.* This is the proof of Theorem 5.8 in the main paper. We prove this theorem with a constructive argument that refines the logic used to prove standard Positive Responsiveness (Proposition 5.5). The proof proceeds by demonstrating that a "tipping-point" profile can always be engineered. The argument follows four steps: 1. We formally define the minimal possible score increase a candidate can receive under CHB. 2. We construct a 'close-race' profile where an arbitrary candidate $p_j$ is losing by a margin smaller than this minimal increase. 3. We apply the single minimal improvement to $p_j$. 4. We show that this minimal change is sufficient to overcome the margin, making $p_j$ the new unique winner.

**1. Define the Minimal Score Change.** For the Constrained Hybrid Borda (CHB) rule, a minimal improvement is defined as a change to a single ballot that increases a candidate's Borda score by exactly one point, without altering its first-place vote count. This distinction is crucial. An improvement that moves a candidate into first place would necessarily increase both its first-place support and its Borda score, resulting in a larger, compound increase to the final Hybrid Score. To isolate the smallest possible quantum of change that the rule can register, we must therefore consider the scenario where a voter raises a candidate's rank but not into the top position. This ensures only the Borda component of the Hybrid Score is affected, representing the true minimal positive change.

A minimal $p_j$*-improvement* that does not alter first-place votes is to have one voter swap the positions of $p_j$ and an adjacent, higher-ranked candidate (who is not the top winner). This increases the total Borda score $B(p_j)$ by exactly 1. However the corresponding increase in the Hybrid Score is a fixed positive amount, which we define as the minimal quantum of change:

$$\Delta H_{\min} = (1 - \lambda) \cdot \left( \frac{1}{k \cdot (m-1)} \right),$$

where let $\lambda < 1$, so $\Delta H_{\min} > 0$.

**2. Construct a 'close race' profile.** We construct an initial preference profile, $\Pi_k$, where an arbitrary candidate $p_j$ is an eligible candidate but narrowly loses to another eligible candidate, $p_q$. We arrange the $k$ ballots with precision to ensure the following conditions hold in profile $\Pi_k$: $(i)$ both $p_j$ and $p_q$ are in the set of eligible candidates, $\mathcal{P}_{\text{eligible}}$ (meaning they both satisfy the $\alpha$-popularity and $\beta$-consensus thresholds), $(ii)$ the Hybrid Score of $p_q$ is greater than the Hybrid Score of $p_j$, so $p_q$ is the winner: $H(p_q) > H(p_j)$, $(iii)$ the winning margin is constructed to be infinitesimally small: we set up the Borda scores and first-place votes such that the difference, $H(p_q) - H(p_j)$, is a positive value that is strictly less than the minimal quantum of change, $\Delta H_{\min}$. Since we can manipulate the scores with single-point precision, creating such a profile is always possible.

**3. Apply the Minimal Improvement.** We create a new profile, $\Pi'_k$, by applying the single minimal improvement to $p_j$. One voter swaps $p_j$ with the candidate ranked just above it. This has the following effects: $(i)$ the Borda score of $p_j$ increases by exactly 1, $(ii)$ the Hybrid Score of $p_j$ increases by exactly $\Delta H_{\min}$, $(iii)$ the new score is $H'(p_j) = H(p_j) + \Delta H_{\min}$, $(iv)$ the scores for the previous winner, $p_q$, cannot increase, so $H'(p_q) \leq H(p_q)$.

**4. The Winner Flips.** We compare the new Hybrid Scores in profile $\Pi'_k$ to show that $p_j$ is now the winner. From our construction, $\Delta H_{\min}$ is greater than the original winning margin, $H(p_q) - H(p_j)$. By rearranging, we get: $H(p_j) + \Delta H_{\min} > H(p_q)$. Since the new score for $p_j$ is $H'(p_j) = H(p_j) + \Delta H_{\min}$, we can substitute to get $H'(p_j) > H(p_q)$. Since the score for $p_q$ did not increase ($H(p_q) \geq H'(p_q)$), it follows that $H'(p_j) > H'(p_q)$. Thus, in the new profile $\Pi'_k$, $p_j$ has the highest Hybrid Score and becomes the winner.

We have shown that a minimal, **single-point** change in the Borda score was sufficient to flip the outcome, proving that the CHB rule is fine-grained responsive. □

# B Proofs for Section 6: Convergence

This part of the appendix provides the full proofs for the lemmas and theorems presented in Section 6 (Convergence) of the main paper.

## B.1 Formal Model of the Stochastic Process

The following proposition is omitted from the main paper.

**Proposition B.1.1** (Time-Homogeneous Markov Chain)**.** The stochastic process $\{S_t\}_{t\geq 0}$ is a time-homogeneous Markov chain.

*Proof.* To establish that our system, represented by the stochastic process $\{S_t\}_{t\geq 0}$, is a time-homogeneous Markov chain, we demonstrate that the transition probabilities from a state $S_t$ to a subsequent state $S_{t+1}$ depend solely on $S_t$. To do this, it is enough to show that the mechanism, for transitioning from any non-terminal state $S_t$ to the next state $S_{t+1}$, is determined entirely by information contained within the state $S_t$:

1. **State-Dependent Components:** The set of unlocked voters, $U(t)$, from which a voter $u$ is selected, is entirely determined by the current state $S_t$.
2. **Static Transition Rules:** The broader environment for state transitions is fixed: $i.$ the overall set of voters $N$ and their preferences $\Pi$ are static, $ii.$ the random sampling process, which involves drawing $k$-samples from $N$, follows a fixed procedure, and $iii.$ while the set of locked voters (part of $S_t$) influences the outcome of this procedure (via the `CHB` function), the underlying rules governing the transition do not change over time.

The sequence of states leading to $S_t$ provides no additional information relevant to the selection of $u$ or the execution of the random sampling. All necessary information is encapsulated in the current state $S_t$. Therefore, the Markov property holds. Since the transition mechanism itself does not depend on the time stamp $t$, the process is *time-homogeneous*. □

## B.2 Proofs of Core Lemmas

**Lemma 6.1.** (Positive Decision Probability). For any non-terminal system state $S_t$ (where the set of unlocked voters $U(t)$ is not empty), the probability that a randomly chosen unlocked voter $u \in U(t)$ transitions to a `LOCK` state is strictly greater than zero. Formally,

$$\mathbf{P}(s_u(t+1) \in \mathcal{P} \mid s_u(t) = \perp, S_t) > 0.$$

Equivalently,

$$\mathbf{P}(s_u(t+1) = \perp \mid s_u(t) = \perp) < 1.$$

*Proof.* To prove this, we need to show the existence of at least one sequence of events with a non-zero probability that forces a `LOCK` outcome. We construct this as follows.

**Define a Specific Sampling Outcome:** Let $u$ be the unlocked voter chosen to perform its rank update. The total population of voters available for sampling is $N' = N \setminus \{u\}$, with size $n-1$. We fix a specific, arbitrary sample of $k$ voters, $K \subset N'$. The probability of drawing this exact sample $K$ in a single round is:

$$\mathbf{P}_K = \binom{n-1}{k}^{-1},$$

where $\binom{n-1}{k}$ is the total number of possible outcomes. Since $k \leq n-1$, this probability is well-defined and strictly positive ($\mathbf{P}_K > 0$).

**Construct a Deterministic Path to LOCK:** Consider the event where voter $u$ draws the exact same sample $K$ for $\tau_{\max}$ consecutive rounds, where $\tau_{\max} = \lfloor \gamma/2 \rfloor + 1$ is the early exit threshold. The probability of this specific sequence of $\tau_{\max}$ identical draws occurring is:

$$\begin{aligned}\mathbf{P}(\texttt{repeated draws}) &= (\mathbf{P}_K)^{\tau_{\max}} \\ &= \left(\binom{n-1}{k}^{-1}\right)^{\tau_{\max}}.\end{aligned}$$

Since $\mathbf{P}_K > 0$, this probability is also strictly positive.

**Analyse the CHB Outcome:** The aggregation rule $F_{CHB}$ used by the protocol must satisfy Resoluteness, which guarantees that for any specific input sample $K$, the output is always a single, well-defined candidate $p^{\text{CHB}}$. In our constructed scenario, every time the voter feeds the fixed sample $K$ into the CHB rule, the resulting outcome is identical. Resoluteness ensures that this identical outcome is always a single candidate, thus preventing the possibility of a tied set of winners. Let this single winner be $p^{\text{CHB}}$, whose counter $C[p^{\text{CHB}}]$ can be incremented.

**Trigger the Early Exit Condition:** By our construction, after $\tau_{\max}$ rounds, the process for voter $u$ is as follows:

- ROUND 1: Sample is $K$, winner is $p^{\text{CHB}}$ and the robustness counter $C[p^{\text{CHB}}]$ becomes 1.
- ROUND 2: Sample is $K$, winner is $p^{\text{CHB}}$ and the robustness counter $C[p^{\text{CHB}}]$ becomes 2.
- ...
- ROUND $\tau_{\max}$: Sample is $K$, winner is $p^{\text{CHB}}$ and the robustness counter $C[p^{\text{CHB}}]$ becomes $\tau_{\max}$.

At this point, the algorithm checks the early exit condition: If $C[p^{\text{CHB}}] \geq \tau_{\max}$. Since the counter is exactly $\tau_{\max}$, the condition is met. The algorithm immediately terminates and returns $\texttt{LOCK}(p^{\text{CHB}})$.

We have constructed a specific sequence of random events (drawing sample $K$ for $\tau_{\max}$ consecutive rounds) that guarantees a `LOCK` outcome. The probability of this sequence, $(p_K)^{\tau_{\max}}$, is strictly greater than zero. Since the total probability of achieving a `LOCK` is the sum of probabilities of all the different (mutually exclusive) sequences leading to a `LOCK`, and we have identified at least one such sequence with a positive probability, the total probability must be strictly positive,

$$\begin{aligned}&\mathbf{P}(\texttt{LOCK}) = \\ &= \mathbf{P}(\texttt{repeated draws}) + \mathbf{P}(\text{sequence 1}) + \\ &\quad + \mathbf{P}(\text{sequence 2}) + \ldots \geq \mathbf{P}(\texttt{repeated draws}) > 0,\end{aligned}$$

which completes the proof. The possibility of a `NO-LOCK` requires the sampled preferences to be so pathologically varied across the $\gamma$ rounds that no winner can gain sufficient traction. For example, when the results are so contradictory that the algorithm cannot build confidence in any single winner. While it is possible for the system to remain undecided, this is not a certainty (due to the fact that there is always a non-zero chance that the random samples align to produce a decisive `LOCK`),

$$\mathbf{P}(\texttt{NO-LOCK}) = 1 - \mathbf{P}(\texttt{LOCK}) < 1.$$

□

**Lemma 6.2.** (Amplification of Plurality). Given two[2] candidates, $p_a$ and $p_b$, in a system state $S_t$ where the number of voters locked on each candidate is $N_a(t)$ and $N_b(t)$ respectively, if $N_a(t) > N_b(t)$, then the probability that a randomly selected unlocked voter $u$ subsequently locks on $p_a$ is strictly greater than the probability $u$ locks on $p_b$.

*Proof.* Let $L_a$ and $L_b$ be the events that voter $u$ locks on candidate $p_a$ and $p_b$, respectively. We show that if $N_a(t) > N_b(t)$, then the probability of locking on the plurality leader is strictly greater, i.e., $\mathbf{P}(L_a \mid S_t) > \mathbf{P}(L_b \mid S_t)$.

The decision of an unlocked voter is based on a random $k$-sample from the electorate. Since the number of voters locked on $p_a$ is greater than for $p_b$, a single voter drawn randomly from the population is more likely to be one locked on $p_a$. This sampling advantage accumulates. Let the random variables $C_a$ and $C_b$ represent the number of voters locked on $p_a$ and $p_b$ within a given $k$-sample. The probability distribution of the count $C_a$ stochastically dominates that of $C_b$. This means that for any threshold $x$, the probability of observing at least $x$ votes for $p_a$ in a sample is greater than or equal to the probability of observing at least $x$ votes for $p_b$. Formally, for all $x \in \{1, \ldots, k\}$, we have $\mathbf{P}(C_a \geq x) \geq \mathbf{P}(C_b \geq x)$, with the inequality being strict for at least some $x$.

The crucial link between this input bias and the output probability is the **Positive Responsiveness** axiom, which any compatible aggregation rule $F$ must satisfy. This axiom has two components that work in tandem:

1. **Monotonicity** acts as a safeguard, ensuring that the greater stochastic support for $p_a$ cannot harm its chances of winning.

2. **Responsiveness** ensures the rule is sensitive to changes in support. It guarantees that an increase in support can be decisive, turning a potential loss or tie into a win. This property is what translates the strict stochastic advantage $p_a$ holds in the input samples into a strict, non-zero probabilistic advantage in the output of the aggregation rule.

Therefore, the combination of monotonicity (preventing harm) and responsiveness (forcing positive action) guarantees that the probability of the aggregation rule selecting $p_a$ in any given round is strictly greater than the probability for selecting $p_b$. Since a voter's final `LOCK` decision is based on tallying the winners from multiple rounds, this per-round advantage carries over to the final outcome. This proves the lemma:

$$\begin{aligned}&\underbrace{\mathbf{P}(C_a \geq x) \geq \mathbf{P}(C_b \geq x)}_{\text{Stochastic Dominance of Input}} \\ \implies &\underbrace{\mathbf{P}(F_{\text{CHB}}(\Pi_k) = p_a) > \mathbf{P}(F_{\text{CHB}}(\Pi_k) = p_b)}_{\text{Result of Positively Responsive Rule}} \\ \implies &\underbrace{\mathbf{P}(L_a \mid S_t) > \mathbf{P}(L_b \mid S_t)}_{\text{Final Lock Probability}}.\end{aligned}$$

□

[2]The logic holds when extended to $m$ candidates however, for simplicity, we restrict the statement (and proof) of lemma to two candidates. The purpose of the two-candidate proof is to isolate and demonstrate the core mechanism: how a difference in existing support between any two candidates translates into a probabilistic advantage for the more popular one.

## B.3 Proof of the Main Convergence Theorem

**Theorem 6.3.** (Almost-Sure Single-Winner Convergence). The Snowveil framework for finding a single winner, when instantiated with any rule satisfying the Core Axioms, is a time-homogeneous Markov chain that almost surely converges in finite time to an absorbing state where a single winner has reached the required quorum $Q$.

*Proof.* The proof proceeds by defining a potential function and proving a series of lemmas about its properties and the dynamics of the underlying Markov chain.

**Lemma B.3.1** (Bounded Potential)**.** The potential function $\Phi(S_t)$ is strictly bounded for all $t$ as follows,

$$0 \leq \Phi(S_t) \leq n^2,$$

where $n$ is the number of all voters.

*Proof.* The lower bound occurs when no voters are locked ($N_j = 0$ for all $j$). The upper bound occurs when all $n$ voters are locked on a single candidate, say $N_1$, giving $\Phi = N_1^2 + N_2^2 + \ldots + N_m^2 = n^2 + \sum_{j:j\neq 1}^{m} 0^2 = n^2$. □

**Lemma B.3.2** (Monotonic Progress)**.** If a voter locks on a candidate $p_k$ during the transition from state $S_t$ to $S_{t+1}$, the given potential function strictly increases and the difference is given by

$$\Delta\Phi = \Phi(S_{t+1}) - \Phi(S_t) = 2N_k(t) + 1.$$

*Proof.* The state change only affects $N_k$, which becomes $N_k + 1$ (in the next state). The new potential is

$$\begin{aligned} \Phi(S_{t+1}) &= \sum_{j\neq k} N_j(t)^2 + (N_k(t)+1)^2 \\ &= \sum N_j(t)^2 + 2N_k(t) + 1 \\ &= \Phi(S_t) + 2N_k(t) + 1. \end{aligned}$$

Since $N_k(t) \geq 0$, the change $\Delta\Phi$ is at least 1. □

**Lemma B.3.3** (Strict Submartingale Property)**.** The sequence of potentials $\{\Phi(S_t)\}$ is a strict submartingale with respect to the process history, for all non-terminal states. That is, the expected potential value at the next state is strictly greater than the potential value of the current state,

$$\mathbb{E}[\Phi(S_{t+1}) \mid S_t] > \Phi(S_t).$$

*Proof.* Let $S_t$ be a non-terminal state of the system. We have

$$\mathbb{E}[\Phi(S_{t+1}) \mid S_t] - \Phi(S_t) = \mathbb{E}[\Delta\Phi \mid S_t],$$

where $\mathbb{E}[\Delta\Phi \mid S_t]$ is the expected change in the potential, which, in turn, equals to the sum of each possible change multiplied by the probability of that change happening. Using the law of total expectation, we have

$$\mathbb{E}[\Delta\Phi \mid S_t] = \sum_{k=1}^{m} \mathbf{P}(s_u(t+1) = p_k \mid S_t) \cdot (2N_k(t) + 1),$$

where $u$ is a voter chosen uniformly at random from $U(t)$, the set of unlocked voters in state $S_t$. By Lemma 6.1, the total probability of locking, $\sum_{k=1}^{m} \mathbf{P}(s_u(t+1) = p_k \mid S_t)$, is greater than 0. By Lemma B.3.2, the potential change for any LOCK event is $\geq 1$. Since all terms in the sum are non-negative and at least one term is strictly positive, the total expected change is strictly positive. Thus, $\mathbb{E}[\Phi(S_{t+1})] > \Phi(S_t)$. □

**Lemma B.3.4** (Instability of Fragmented States)**.** A system state $S_t$ is a fragmented state if it is not an absorbing state and the set of unlocked voters $U(t)$ is not empty. Any such state is a transient state of the Markov chain.

*Proof.* We prove that any system state characterised by fragmented consensus is unstable and that the process inevitably moves away from it.

**Definition B.1** (Fragmented State)**.** A state $S_t$ is a fragmented state when it is not an absorbing state, which means that for all candidates $p_j \in \mathcal{P}$, the number of locked voters is strictly less that the specified quorum, $N_j(t) < \lceil Q \cdot n \rceil$. Furthermore, the set of unlocked voters $U(t)$ is non-empty, so the process has not terminated due to all voters being locked. This corresponds to a non-absorbing, non-terminal state in our defined Markov chain.

**The System Cannot Remain Static:** By Lemma 6.1 (Positive Decision Probability), we know that for any unlocked voter $u \in U(t)$ who is selected to update, there is a strictly positive probability that they will transition to a LOCK state. This implies that the probability of the system state changing is greater than zero, i.e., $\mathbf{P}(S_{t+1} \neq S_t \mid S_t) > 0$. Therefore, the system cannot remain trapped in the exact state $S_t$ indefinitely.

Destabilization via Positive Feedback: The instability of a fragmented state is driven by the positive feedback mechanism established in Lemma 6.2 (Amplification of Plurality). We analyse the system's evolution in two scenarios:

- $(A)$**: An Existing Plurality.** Assume there is a candidate $p_a$ that is the unique plurality leader, such that $N_a(t) > N_j(t)$ for all other candidates $p_j \in \mathcal{P}$. According to Lemma 5, the probability that the next unlocked voter locks on $p_a$ is strictly greater than the probability of them locking on any other candidate. Each time a voter locks on $p_a$, its lead increases, which in turn makes it even more probable that the next voter will also lock on $p_a$. This self-reinforcing dynamic actively drives the system away from the fragmented state and towards achieving a quorum for $p_a$.
- $(B)$**: A Perfectly Balanced State.** Assume there are two leading candidates, $p_a$ and $p_b$, and their support is perfectly balanced (e.g., $N_a(t) = N_b(t)$ and this count is higher than for any other candidate). While the probabilities of locking on $p_a$ or $p_b$ may be equal, the process is stochastic. Due to the randomness of sampling, it is not a certainty that this balance will be maintained. A LOCK event will eventually occur (by Lemma 3), and this event will necessarily break the tie, creating a new state where one candidate has a slight plurality. This new state falls under Scenario A, and the positive feedback loop takes over.

**All Fragmented States are Transient:** In a finite Markov chain, a state is defined as transient if, upon leaving it, there is a non-zero probability of never returning. From any fragmented state $S_t$, we have established that the system will eventually transition to a new state $S_{t+1} \neq S_t$. By Lemma B.3.2 (Monotonic Progress), any transition to a LOCK state strictly increases the potential function, $\Phi(S_{t+1}) > \Phi(S_t)$. Since the potential function $\Phi$ can never decrease (as voters transition only from unlocked to locked), the system never returns to a state with a lower potential value. Therefore, any fragmented state $S_t$ is a transient state. By the fundamental properties of finite Markov chains, a process cannot remain in transient states indefinitely and must, with probability 1, eventually reach a recurrent class (a state where it cannot escape from). In our model, the only recurrent states are the absorbing states (where a quorum is met). This proves that any state of fragmented consensus is inherently unstable. The system is designed to break such ties and almost surely evolves away from fragmentation towards an absorbing state. □

**Conclusion of the proof.** The Snowveil process is a finite-state, time-homogeneous Markov chain. The sequence of potentials $\{\Phi(S_t)\}$ is a bounded submartingale and therefore must converge almost surely by the Martingale Convergence Theorem. Since all non-absorbing states are transient, the process cannot converge to a non-absorbing state. Thus, it must almost surely enter an absorbing state in finite time. This completes the proof of Theorem 6.3. □

# C Proof for Section 7: Strategic Resilience

**Proposition 7.1.** (Minimal Coalition Size). To neutralise the probabilistic advantage of an truthful Borda winner $p^*$ with a per-voter margin of $\delta$, a coordinated manipulating coalition requires a minimum size $c$ such that $c \geq n \cdot (\delta/(\delta + m - 1))$, which is $\Omega(n)$.

*Proof.* Recall that $p^*$ is the canonical outcome, i.e. $p^* = F_{\text{CHB}}(\Pi)$ . In this section, we analyse the resilience of $p^*$ against burying attacks by focusing on the Borda component of the CHB rule. The Borda component represents the protocol's primary signal for broad consensus.

Consider a set of $n$ voters that consists of $n - c$ truthful voters and a coalition of $c$ manipulators. The truthful winner (the candidate who would win if all ballots were truthful) is $p^*$, however the coalition's preferred candidate is $p_c$. We analyse the Borda score, as it is the core component of CHB and is mathematically tractable. Let $B(p, R_i)$ be the Borda score (from 0 to $m-1$) for candidate $p$ on voter $i$'s ballot $R_i$.

We assume that the $c$ manipulators' coalition employ an optimal strategy: they rank $p_c$ first (score $m - 1$) and $p^*$ last (score 0), generating a per-voter malicious margin of $-(m - 1)$, which is the largest possible negative margin, or, 'anti-signal'. A voter's LOCK decision is based on the results from a random $k$-sample. For the manipulation to have a chance, the expected Borda score difference $\mathbb{E}[\Delta_k]$ in a random $k$-sample must be non-positive. This expectation is a weighted average of the truthful and malicious margins over the $k$ samples: let $\Delta_k$ be the random variable for the total Borda score difference $(B(p^*) - B(p_c))$ within a $k$-sample. We calculate $\mathbb{E}[\Delta_k]$ and solve for $c$. By linearity of expectation, the expected total difference in a sample of $k$ is $k$ multiplied with the expected difference from a single voter drawn randomly from the entire population,

$$\mathbb{E}[\Delta_k] = k \cdot \mathbb{E}[B(p^*, R_{\text{rand}}) - B(p_c, R_{\text{rand}})].$$

The expected difference from a single random voter is a weighted average of the truthful and malicious margins:

$$\begin{aligned}\mathbb{E}_{\text{rand}} = &\ \mathbf{P}(\text{voter is truthful}) \cdot \delta + \\ &+ \mathbf{P}(\text{voter is malicious}) \cdot (-(m-1)) = \\ &= \frac{n-c}{n} \cdot \delta + \frac{c}{n} \cdot (-(m-1)).\end{aligned}$$

The manipulator's objective is $\mathbb{E}[\Delta_k] \leq 0$. Applying this, we get

$$\mathbb{E}[\Delta_k] = k \cdot \mathbb{E}_{\text{rand}} = k \cdot \left[\frac{n-c}{n} \cdot \delta - \frac{c}{n} \cdot (m-1)\right] \leq 0.$$

Since $k > 0$ and $n > 0$, we simplify by multiplying out the denominators,

$$\begin{aligned}&(n-c) \cdot \delta - c \cdot (m-1) \leq 0 \Leftrightarrow \\ &n \cdot \delta - c \cdot \delta \leq c \cdot (m-1) \Leftrightarrow \\ &n \cdot \delta \leq c \cdot (m-1) + c \cdot \delta \Leftrightarrow \\ &n \cdot \delta \leq c \cdot (m - 1 + \delta).\end{aligned}$$

Solving for $c$, we get

$$c \geq n \cdot \frac{\delta}{\delta + m - 1}.$$

Since $\delta$ and $m$ are constants independent of the electorate size $n$, the required coalition size is a linear fraction of the electorate, thus $c = \Omega(n)$. □

# D Proofs for Section 8: Scalability

**Lemma 8.1.** (Positive $n$-Independent Lock Probability) [Positive $n$-Independent Lock Probability] For any non-terminal system state, the probability that a randomly chosen unlocked voter transitions to a LOCK state on the canonical outcome $p^*$ is lower-bounded by a positive constant $c_2 > 0$, which is independent of the number of voters.

*Proof.* A formal proof of this lemma requires the use of concentration bounds. Our protocol uses sampling without replacement from a finite population, for which the tightest bounds are given by results such as Serfling's inequality. However, in the common and practical regime where the sample size $k$ is significantly smaller than the total population $n$ ($k \ll n$), the statistical difference between sampling with and without replacement is negligible. We therefore model the sampling process as drawing $k$ independent and identically distributed (i.i.d.) variables. This standard approximation allows for the use of the simpler and more established Chernoff-Hoeffding bound, providing a clear and analytically tractable lower bound on the lock probability.

The proof is split into five subsections, **D.1 - D.5**, for clarity. The goal of this proof is to demonstrate that the probability of a single voter successfully locking on the correct global winner, $p^*$, in a single execution of the `UpdateVoter` protocol is lower-bounded by a positive constant, $c_2$, that is independent of the total number of voters, $n$. The argument proceeds in four steps. First, we frame the problem by defining the single-round success probability for a voter. Second, we use the Chernoff-Hoeffding bound to show that the scores within a random $k$-sample are a high-fidelity proxy of the global scores. Third, we use the union bound to aggregate the probability of failure against all competing candidates in a single round. Finally, we formalise the resulting $n$-independent constant and connect it to the overall lock probability.

## D.1 The Local Decision as a Probabilistic Event

We begin the proof by analysing the protocol's fundamental action: the (local) decision process of a single, unlocked voter $v$. The voter samples $k$ other voters uniformly at random and collects their individual preference rankings, $\{R_1, R_2, \ldots, R_k\}$. This multiset of $k$ rankings constitutes the preference profile that we denote as $\Pi_k$.

While the aggregation rule $F_{\text{CHB}}$ is a deterministic function, its output, which we denote as the random variable

$$p_{\text{local}} = F_{\text{CHB}}(\Pi_k),$$

is probabilistic because its input $\Pi_k$ is the result of a random draw. The random variable $p_{\text{local}}$ is the local winner of this specific round by applying the CHB rule to this specific sample $\Pi_k$. Thus $F_{\text{CHB}}$ denotes the aggregation function applied to a single sampling iteration. The final `LOCK` decision is determined by the `Voter Update Process`, which aggregates these $p_{\text{local}}$ outcomes over $\gamma$ (robustness) rounds.

The center of the proof relies on the probability of the event $\{p_{\text{local}} = p^*\}$, where $p^*$ is the correct global winner (canonical outcome) as defined in the system model. We show that the set of numerical conditions that cause $p^*$ to win under the CHB rule in the global profile $\Pi$ also hold simultaneously within a random $k$-sample with high $n$-independent probability.

The immediate objective is to prove that the probability $\mathbf{P}(p_{\text{local}} = p^*)$ is lower-bounded by a positive constant $c_1$ that is independent of the total network size $n$. This $c_1$ then serves as the foundation for proving that the overall `LOCK` probability $\mathbf{P}_{\text{LOCK}}$ is also lower-bounded by another constant $c_2$. We give the definition of the overall $\mathbf{P}_{\text{LOCK}}$ probability at the end of this section.

Formally, we need to show that

$$\exists c_1 > 0 \text{ such that } \mathbf{P}(p_{\text{local}} = p^*) \geq c_1, \forall n > k.$$

Intuitively, this expresses that no matter how unfavourable the circumstances are, the probability of a voter making the correct decision in a given single round never falls below the positive constant $c_1$.

This fixed constant $c_1$ acts as a floor for the probability of success ($\mathbf{P}(p_{\text{local}} = p^*)$), and this floor holds true for any network size $n$, as long as $n$ is big enough to sample from. Proving this claim is the cornerstone of the protocol's $n$-independent scalability, as it establishes that a single voter's ability to make a correct decision does not degrade as the network grows.

**Definition D.1** (Overall `LOCK` Probability ($\mathbf{P}_{\text{LOCK}}$))**.** We define $\mathbf{P}_{\text{LOCK}}$ as the probability that an unlocked agent transitions to a `LOCK` state during a single execution of the `Update Voter Process` (Algorithm 1). Since the single round success probability is lower-bounded by $c_1$, the overall `LOCK` probability is lower-bounded by

$$\mathbf{P}_{\text{LOCK}} \geq \sum_{i=\tau_{\max}}^{\gamma} \binom{\gamma}{i} \cdot c_1^i \cdot (1 - c_1)^{\gamma - i} = c_2.$$

where recall that $\gamma$ is the total number of sampling rounds an agent performs, $\tau_{\max}$ the threshold number of wins required to lock, $c_1$ is the minimum probability of winning a single round, while $\binom{\gamma}{i}$ expresses the number of different ways to choose $i$ winning rounds out of $\gamma$.

Remark on Definition D.1: The `Voter Update Process` (Algorithm 1) terminates early as soon as the counter reaches the threshold $\tau_{\max}$. We note that this event is mathematically equivalent to the condition where a (full) sequence of $\gamma$ rounds contains at least $\tau_{\max}$ successes. Therefore, to compute the probability $\mathbf{P}_{\text{LOCK}}$, it is valid to calculate the tail of the binomial distribution (sum from $\tau_{\max}$ to $\gamma$).

## D.2 Bounding Sample Deviations via Concentration Bounds

This step formally bridges the gap between the global preference profile $\Pi$ and the local information available in a random $k$-sample. The CHB rule's outcome depends on Borda scores and first-place vote counts. Therefore we need to show that a random sample is a high-fidelity proxy for both of these metrics. We define the per-voter random variables for each score, establish their global margins, define the 'bad events' where a sample deviates misleadingly, and apply concentration bounds to show that the probability of these bad events occurring is exponentially small in the sample size $k$.

### Defining Per-Voter Random Variables and Global Margins

We define two types of random variables based on drawing a single voter from the population, as follows.

**Borda Score Variables:**

**Definition D.2** (Score Difference)**.** For each of the $m - 1$ competing candidates $p_j$ (where $p_j \neq p^*$), we define a random[3] variable, $D_j$, which represents the score difference between $p^*$ and $p_j$ for a single, randomly drawn voter from the entire pool of voters (the global population) as

$$D_j = \text{Score}(p^*) - \text{Score}(p_j).$$

[3] This is a random variable because each voter has their own opinion and thus their own score difference $D_j$. Their specific value of $D_j$ is unknown until they are randomly queried.

**Definition D.3** (Global mean of Score Difference)**.** For each competitor $p_j$, the true global mean $\delta_j$ is the average of the score difference $D_j$ calculated across every single voter in the entire population. Formally, we have

$$\delta_j = \mathbb{E}[D_j] = \frac{1}{n} \cdot \sum_{i=1}^{n} D_{j,i},$$

where $D_{j,i}$ is the individual $D_j$ value of a voter $i$ in the global population.

**Interpretation of the Global mean $\delta_j$ as the Per-Voter Margin of Victory.** We connect this mean to the idea of an election victory: if $\delta_j > 0$, then $\mathbb{E}[D_j] > 0$, meaning that, on average, a voter in the electorate gives a higher score to candidate $p^*$ than to candidate $p_j$. Therefore, $\delta_j$ is the **per-voter margin of victory**, as it quantifies the average amount by which a single voter favours $p^*$ over $p_j$. For instance, let $\delta_j = 0.75$. This means that, on average, any randomly chosen voter rates $p^*$ three-quarters of a point higher than they rate $p_j$. It is the fundamental measure of $p^*$'s popularity advantage over $p_j$ across the whole electorate.

**Plurality Count Variables:** For each candidate $p_j$, let $Y_j$ be an indicator variable that equals to 1 if a random voter ranks $p_j$ first, and 0 otherwise. The global mean is the global first-place proportion, $\mu_{Y_j} = \mathbb{E}[Y_j] = T_{\text{glob}}(p_j)/n$. From this, we define the crucial $\alpha$-popularity margin for the winner $p^*$. Assuming $p^*$ is $\alpha$-popular globally i.e., $\mu_{Y_*} > \alpha$, we define this margin as

$$\delta_\alpha = \mu_{Y_*} - \alpha.$$

This value $\delta_\alpha > 0$ represents the 'safety buffer' that $p^*$'s global first-place support has over the required threshold.

### Defining 'Bad Sample' Events

A 'bad sample' is one that misrepresents the global profile in a way that could alter the CHB outcome. We define the primary bad events (for the case where $p^*$ is a popular Borda winner):

- **Bad event $E_j$ (Borda Reversal):** For a competitor $p_j$, this is the event where the Borda scores in the sample favour $p_j$. Let $S_{D,k}(j)$ be the sum of $k$ draws of the random variable $D_j = \text{Score}(p^*) - \text{Score}(p_j)$. Then the bad event $E_j$ is the event $S_{D,k}(j) \leq 0$.
- **Bad event $B$ (Loss of $\alpha$-Popularity):** This is the event where the winner $p^*$ is not $\alpha$-popular in the sample. Let $T_{\text{samp}}(p^*) = \sum_{i=1}^{k} Y_{*,i}$ be the first-place count for $p^*$ in the $k$-sample. Then $B$ is the bad event $T_{\text{samp}}(p^*) < \lceil \alpha \cdot k \rceil$.

## D.3 Bounding the Probabilities of Bad Events

We now apply the Chernoff-Hoeffding bound to find an upper limit on the probability of each bad event discussed in previous section.

### Bounding the Probability of Borda Reversal $\mathbf{P}(E_j)$

We first present the definitions used in the proof of bounding $\mathbf{P}(E_j)$.

**Definition D.4** (Sample Sum $S_{D,k}$)**.** The total score difference in a sample of $k$ voters is the sum of $k$ individual difference variables (independent and identically distributed instances of variable $D_j$)

$$S_{D,k} = D_{j,1} + \cdots + D_{j,k},$$

where $D_{j,i}$ is the score difference for the $i$-th voter in the sample.

**Definition D.5** (Expected Sample Sum $\mu$)**.** The expected sample sum $\mu$ is the average outcome that we expect for $S_{D,k}$, this is

$$\begin{aligned} \mu = \mathbb{E}[S_{D,k}] &= \mathbb{E}[D_{j,1} + \cdots + D_{j,k}] \\ &= \mathbb{E}[D_{j,1}] + \mathbb{E}[D_{j,2}] + \ldots + \mathbb{E}[D_{j,k}] \\ &= k \cdot \mathbb{E}[D_j] = k \cdot \delta_j. \end{aligned}$$

Since $p^*$ is the true winner, then $\delta_j > 0$, thus, in turn, $\mu > 0$.

We proceed with our goal which is to find an upper bound for the probability of the failure event $E_j$, $\mathbf{P}(S_{D,k} \leq 0)$, using these definitions.

### Applying the Chernoff-Hoeffding Bound to the Difference

We apply the **Chernoff-Hoeffding bound** to find the probability of this failure event, $E_j$. This bound is a one-sided version of Hoeffding's Inequality, which gives an upper bound on the chance that the sum of several random variables is significantly lower than its expected value.

**Definition D.6** (Chernoff-Hoeffding bound)**.** Let $X_1, X_2, \ldots, X_n$ be a set of $n$ independent random variables, that are bounded, meaning each $X_i$ is almost surely confined to an interval $[a_i, b_i]$. The length of this interval, $b_i - a_i$, is the range of the variable. Let $S_n$ be the sum of these variables, $S_n = \sum_{i=1}^{n} X_i$, and $\mathbb{E}[S_n]$ the expected value (or mean) of this sum. Then, for any positive value $t > 0$, the probability that the sum $S_n$ deviates below its expected value by at least $t$ is bounded as

$$\mathbf{P}\left(S_n - \mathbb{E}[S_n] \leq -t\right) \leq \exp\left(-\frac{2t^2}{\sum_{i=1}^{n}(b_i - a_i)^2}\right).$$

### The Hoeffding Bound Transformation

We continue by making the failure event, $S_{D,k} \leq 0$, fit this bound structure. We have

$$\begin{aligned} S_{D,k} \leq 0 &\Leftrightarrow S_{D,k} - \mu \leq 0 - \mu \\ &\Leftrightarrow S_{D,k} - \mu \leq -\mu, \end{aligned}$$

which matches the Hoeffding structure, $\text{Sum} - \mathbb{E}[\text{Sum}] \leq -t$, for $\text{Sum} = S_{D,k}$, $\mathbb{E}[\text{Sum}] = \mathbb{E}[S_{D,k}] = \mu$ and $t = \mu$.

**Applying the Bound**

With $t = \mu$, we plug everything into the Hoeffding formula to find the probability of failure $P(E_j)$. First, we establish the form of the event,

$$\mathbf{P}(E_j) = \mathbf{P}(S_{D,k} \leq 0) = \mathbf{P}(S_{D,k} - \mu \leq -\mu).$$

Applying the Hoeffding bound, as in Definition D.6, where $S_{D,k} = \sum_{i=1}^{k} D_{j,i}$ gives,

$$\mathbf{P}(E_j) \leq \exp\left(-\frac{2\mu^2}{\sum_{i=1}^{k}(\text{range of } D_{j,i})^2}\right). \tag{1}$$

For Hoeffding's inequality to apply, the random variables in the sum must satisfy two crucial conditions: *independence* - justified by the nature of uniform random sampling from the electorate (see Appendix for an illustration) - and *boundedness*. We have discussed their independence and we continue to show that they are also bounded.

**Defining the Range for CHB**

Recall that the score difference is given by $D_j = \text{Score}(p^*) - \text{Score}(p_j)$ for a single voter. The CHB rule is a hybrid rule, however its range for the purpose of the Chernoff-Hoeffding bound is ultimately determined by the widest possible score difference. To find this, we analyse the three types of scores that CHB uses for a candidate $p_j$:

- **Plurality Score $t_j$:** A voter gives a score of 1 if $p_j$ is their top choice, and 0 otherwise. The range of this score is $[0, 1]$.
- **Borda Score $B(p_j)$:** A voter gives a score from 0 to $m - 1$. The range of this score is $[0, m - 1]$.
- **Hybrid Score $H(p_j)$:** This is a weighted average of the normalised Borda and Plurality scores. Since both underlying scores are bounded, the Hybrid score is also bounded within $[0, 1]$.

**Determining the Maximum Range of the Difference**

To find the most conservative (i.e., widest) possible range, we look at the component with the largest score variation, which we can see from previous step that this is the Borda score. We have:

- The score for $p^*$ can be as high as $m - 1$ and as low as 0.
- The score for $p_j$ can also be as high as $m - 1$ and as low as 0.
- The maximum possible score difference occurs if a voter ranks $p^*$ first and $p_j$ last, giving $D_j = (m - 1) - 0 = m - 1$.
- The minimum possible difference occurs if they rank $p_j$ first and $p^*$ last, giving $D_j = 0 - (m - 1) = -(m - 1)$.

Therefore, the variable $D_j$ is strictly bounded within the interval $[-(m - 1), m - 1]$, thus the total range is $(m - 1) - (-(m - 1)) = 2 \cdot (m - 1)$. Substituting this, together with $\mu = k \cdot \delta_j$ into inequality (1), the denominator becomes $k \cdot (2(m - 1))^2$ and gives a final bound of

$$\begin{aligned}\mathbf{P}(E_j) &\leq \exp\left(-\frac{2 \cdot (k \cdot \delta_j)^2}{k \cdot (2 \cdot (m-1))^2}\right)\\ &= \exp\left(-\frac{2 \cdot k^2 \cdot \delta_j^2}{4 \cdot k \cdot (m-1)^2}\right)\\ &= \exp\left(-\frac{k \cdot \delta_j^2}{2 \cdot (m-1)^2}\right).\end{aligned}$$

This final expression shows that the probability of the sample failing to reflect $p^*$'s victory over any *single* competitor drops exponentially as the sample size $k$ increases. Because this guarantee of sample fidelity holds - that is, the probability of any score being significantly misleading is negligible - the relative ordering of scores and the winner's margin of victory are preserved with high probability.

**Bounding the Probability of Losing $\alpha$-Popularity $\mathbf{P}(B)$**

The random variable $Y_*$ (first-place indicator) has a range of 1 (from 0 to 1). The sum $T_{\text{samp}}(p^*)$ has an expected value of $\mathbb{E}[T_{\text{samp}}(p^*)] = k \cdot \mu_{Y_*}$. The bad event $B$ is $T_{\text{samp}}(p^*) < \alpha \cdot k$ (we ignore the ceiling function for a simpler bound). We transform this to fit the Hoeffding structure,

$$\begin{aligned}T_{\text{samp}}(p^*) &< \alpha \cdot k\\ T_{\text{samp}}(p^*) - \mathbb{E}[T_{\text{samp}}(p^*)] &< \alpha \cdot k - k \cdot \mu_{Y_*}\\ T_{\text{samp}}(p^*) - \mathbb{E}[T_{\text{samp}}(p^*)] &< -k \cdot (\mu_{Y_*} - \alpha)\\ T_{\text{samp}}(p^*) - \mathbb{E}[T_{\text{samp}}(p^*)] &< -k \cdot \delta_\alpha.\end{aligned}$$

Applying the Chernoff-Hoeffding bound with $t = k \cdot \delta_\alpha$ and a range of 1 for each of the $k$ variables gives:

$$\begin{aligned}\mathbf{P}(B) &\leq \exp\left(-\frac{2 \cdot (k \cdot \delta_\alpha)^2}{\sum_{i=1}^{k} 1^2}\right)\\ &= \exp\left(-\frac{2 \cdot k^2 \cdot \delta_\alpha^2}{k}\right)\\ &= \exp(-2 \cdot k \cdot \delta_\alpha^2).\end{aligned} \tag{2}$$

This shows the probability of the winner $p^*$ wrongly appearing unpopular in the sample also drops exponentially with $k$.

## D.4 Single-Round Failure Probability via Union Bound

The event that the sample is misleading ($E_{\text{failure}}$) is the union of all possible bad events: a Borda reversal against any competitor ($E_j$), or the winner losing its $\alpha$-popularity ($B$). Formally, this event is given by

$$E_{\text{failure}} = B \cup \left(\bigcup_{j \neq *} A_j\right).$$

We then use the union bound to find an upper limit on the total probability of the failure event,

$$\mathbf{P}(E_{\text{failure}}) \leq \mathbf{P}(B) + \sum_{j \neq *} \mathbf{P}(A_j).$$

By substituting the bound derived in inequality (2), we get

$$\mathbf{P}(E_{\text{failure}}) \leq \exp(-2 \cdot k \cdot \delta_\alpha^2) + \sum_{j \neq *} \exp\left(-\frac{k \cdot \delta_j^2}{2 \cdot (m-1)^2}\right). \quad (3)$$

Our goal is to find a single expression that is greater than or equal to every term in the sum above. The function $e^{-x}$ gets larger as its exponent $x$ gets smaller (closer to zero). Therefore, the largest exponential bound comes from the smallest positive coefficient in the exponents. We look at the coefficients of $-k$ in the exponents,

- For the plurality term: $C_\alpha = 2\delta_\alpha^2$.
- For the Borda terms: $C_j = \frac{\delta_j^2}{2(m-1)^2}$.

To find the minimum of all these values, we define a new constant $C_{\min}$ that represents this minimum coefficient among all,

$$C_{\min} = \min\left(2\delta_\alpha^2, \min_{j \neq *}\left(\frac{\delta_j^2}{2(m-1)^2}\right)\right).$$

By definition, $C_{\min}$ is smaller than or equal to any of the individual coefficients meaning that the exponent $-C_{\min} \cdot k$ is larger than or equal to any other exponent. Therefore, $e^{-C_{\min} \cdot k}$ is an upper bound for every single term in the RHS sum of (3). This gives

$$\mathbf{P}(E_{\text{failure}}) \leq m \cdot \exp(-C_{\min} \cdot k).$$

We have converted the sum of different exponentials into a single exponential term multiplied by the number of terms, $m$. The final step is a notational simplification to make the bound's dependence on $k$ and the margins clearer.

Observe that the term $C_{\min}$ is some positive constant that depends on the various $\delta$ margins. Let's define a simplified $\delta_{\min}^2$, as the smallest of all the margin values (after scaling them to be comparable). Our constant $C_{\min}$ is directly proportional to this $\delta_{\min}^2$. Thus we can write

$$C_{\min} = \eta \cdot \delta_{\min}^2 \quad \text{for some positive constant } \eta.$$

Substituting this back into our bound, we get

$$\mathbf{P}(E_{\text{failure}}) \leq m \cdot \exp(-\eta \cdot k \cdot \delta_{\min}^2)$$

In asymptotic notation, a function $f(k)$ is in $\Omega(g(k))$ if it is bounded below by a positive constant multiple of $g(k)$ for large enough $k$. Here, our exponent's magnitude, $\eta \cdot k \cdot \delta_{\min}^2$, is bounded below by a constant multiple of $k \cdot \delta_{\min}^2$. Therefore, we can write

$$\eta \cdot k \cdot \delta_{\min}^2 = \Omega(k \cdot \delta_{\min}^2).$$

Replacing this in the exponent gives us the final, simplified expression

$$\mathbf{P}(E_{\text{failure}}) \leq m \cdot \exp(-\Omega(k \cdot \delta_{\min}^2)),$$

which cleanly communicates that the failure probability drops exponentially with the sample size $k$ and the square of the minimum 'winning margin' $\delta_{\min}$, with the factor $m$ accounting for the number of ways a failure can occur. This formula gives us a concrete upper bound on the probability of our sample being misleading with respect to the full CHB rule.

## D.5 The $n$-Independent Lower Bound

The preceding analysis provides a robust upper bound on the probability of a misleading sample, $\mathbf{P}(E_{\text{failure}})$. We use this upper bound to formalise the lower bound on a voter's single-round success probability, $c_1$. We have that

$$\mathbf{P}(p_{\text{local}} = p^*) = 1 - \mathbf{P}(E_{\text{failure}}) \geq 1 - m \cdot \exp\left(-\Omega(k \cdot \delta_{\min}^2)\right) := c_1.$$

Since the constants hidden by the $\Omega$ notation depend only on the protocol parameters $m$ and the global preference profile margins $\delta_{\min}$, but not on the electorate size $n$, this proves that $c_1$ is a positive constant that is independent of $n$.

Having established this $n$-independent single-round success probability $c_1$, we apply the Definition D.1, in which we have established that the overall lock probability $\mathbf{P}_{\texttt{LOCK}}$ for the full $\gamma$-round `UpdateVoter` process is lower-bounded by the tail of the Binomial distribution parameterised by $c_1$. Therefore, $\mathbf{P}_{\texttt{LOCK}}$ is also lower-bounded by a positive $n$-independent constant $c_2$. This completes the proof of Lemma 8.1. □

**Theorem 8.2.** (Expected O(n) Convergence Time) The Snowveil protocol converges on a single winner in an expected number of steps bounded by $O(n)$.

*Proof.* The proof relies on two foundational results established previously:

1. The protocol is live and is guaranteed to converge to an absorbing state (as established in Section 6 (on Convergence) through the submartingale analysis, see Theorem 6.3).
2. The probability $\mathbf{P}_{\texttt{LOCK}}$ of a single `LOCK` event occurring at any non-terminal step is lower-bounded by an $n$-independent positive constant, $c_2$ (see Lemma 8.1).

We model the convergence as the process of moving voters from an unlocked to a locked state. The protocol terminates when a quorum is reached, which requires at most $n$ successful `LOCK` events in total. This is because there are $n$ voters in the entire system, and each voter can only perform a state change only once (per winner-selection iteration round).

From probability theory, the expected number of steps to achieve a success is $\frac{1}{p}$, where $p$ is the success probability. In our case, this is $\frac{1}{\mathbf{P}_{\texttt{LOCK}}}$: Since the success probability $\mathbf{P}_{\texttt{LOCK}}$ is at least $c_2$, we take the reciprocal of both sides, to get an upper bound on the expected number of steps (or attempts) required to achieve a single success. We have

$$\frac{1}{\mathbf{P}_{\texttt{LOCK}}} \leq \frac{1}{c_2}.$$

Let $C_{\texttt{LOCK}} = 1/c_2$ be this constant expected time per `LOCK` event. The total expected time to convergence, $\mathbb{E}[T]$, is therefore bounded by the maximum number of `LOCK` events required multiplied by the constant expected time per event, i.e.

$$\mathbb{E}[T] \leq (\text{Max LOCK events}) \cdot (\text{Expected steps per LOCK event}) = n \cdot C_{\texttt{LOCK}}.$$

Since $C_{\texttt{LOCK}}$ is a constant independent of $n$, the total expected number of steps for the protocol to converge is $O(n)$. This completes the proof. □

# E Experimental Details

## E.1 Baseline Parameters

Unless specified otherwise in a given experiment, all simulations are run with the baseline parameters detailed in Table 2.

Table 2: Baseline Simulation Parameters

| Parameter | Symbol | Value | Description |
|---|---|---|---|
| Voters | $n$ | 100 | Total voters |
| Candidates | $m$ | 5 | Total candidates |
| Sample Size | $k$ | 10 | Voters per sample |
| Robustness Rounds | $\gamma$ | 10 | Rounds per decision |
| Early Exit | $\tau_{\text{max}}$ | 6 | $\lfloor \gamma/2 \rfloor + 1$ |
| Plurality Threshold | $\tau_{\text{min}}$ | 3 | Min wins for lock |
| Hybrid Weight | $\lambda$ | 0.5 | Borda/Plurality balance |
| Popularity Filter | $\alpha$ | 0.1 | Min first-place support |
| Consensus Filter | $\beta$ | 0.8 | Min Borda score % |
| Global Quorum | $Q$ | 0.67 | Required agreement |

## E.2 Parameter trade-offs

This section defines the role of each parameter and discusses the practical trade-offs involved in their configuration.

**Sample size $k$: Fidelity vs. Efficiency**

The sample size, k, is the most fundamental parameter governing the performance and accuracy of the Snowveil protocol. It directly controls the trade-off between the efficiency of the local decision process and the fidelity of the information a voter gathers. A small $k$ minimises network messages and computational load, but the resulting sample may not accurately reflect the global distribution of preferences. Conversely, a large $k$ provides a high-fidelity snapshot of the electorate's will at the cost of increased overhead.

**Trade-off Analysis**

The choice of $k$ creates a clear and quantifiable set of trade-offs:

- Increasing $k$ directly increases the message overhead of the protocol ($O(k)$ messages per decision round) and the local computation time ($O(k \cdot m)$ for the CHB rule). However, by dramatically improving decision accuracy at the local level, it reduces the likelihood of voters locking on incorrect candidates. This reduction in dissent and contention likely accelerates global convergence, as the system spends less time in fragmented states.
- Decreasing $k$ makes the protocol more lightweight in terms of network load and computation. However, this comes at the cost of increased risk of local sampling errors. These errors can introduce noise and transient support for incorrect candidates, potentially prolonging the time it takes for the entire system to reach a stable, global consensus.

**Aggregation Logic ($\alpha$, $\beta$, $\lambda$): Defining the Winner's Character**

While the parameter $k$ governs the accuracy of the information gathered, the parameters of the Constrained Hybrid Borda (CHB) rule -$\alpha$, $\beta$, and $\lambda$- govern the interpretation of that information. They allow a community to codify its own social values and define what it considers a "good" or "legitimate" outcome. The guidance for setting these parameters is therefore less about optimisation and more about choices regarding normative trade-offs.

**Guidance for $\lambda$: The Consensus-Plurality Balance**

The hybrid weight $\lambda$ is the most direct tool for navigating the classic social choice dilemma between broad consensus and strong, passionate support. It directly controls the final score calculation for eligible candidates, as the Hybrid Score $H(p_j)$ is a convex combination controlled by $\lambda$,

$$H(p_j) = (1-\lambda)\cdot(\text{Norm. Borda Score})+\lambda\cdot(\text{Norm. Plurality}).$$

The choice of $\lambda$ is a community's answer to the question: "Do we prefer a winner everyone can live with, or a winner a large group is passionate about?"

- $\lambda = 0$ **(Pure Consensus):** The rule becomes a pure Borda count among eligible candidates. This favours candidates who are ranked reasonably well by many voters, even if they are the first choice of few. It produces the least controversial winner but risks selecting a "compromise candidate" that inspires no one.
- $\lambda = 1$ **(Pure Plurality):** The rule effectively becomes a First-Past-the-Post election among eligible candidates. This favors candidates with a strong, dedicated base of support, even if they are ranked very low by others. It produces a winner with a strong mandate but risks selecting a divisive candidate who is strongly disliked by the majority.
- $0 < \lambda < 1$ **(Hybrid Approach):** A choice like $\lambda = 0.5$ seeks a balance, giving equal weight to both broad consensus and passionate support.

**Guidance for $\alpha$: The Popularity Filter**

The $\alpha$ parameter acts as a "barrier to entry" for the CHB rule's hybrid mechanism. It ensures that any candidate considered must have a minimum level of first-place support, $\alpha$-popular. This is a prerequisite for a candidate to win via the Borda-only path (Step 2 of the CHB rule) or to become eligible for the hybrid calculation (Step 3). The choice of $\alpha$ is a trade-off between inclusivity and legitimacy. It answers the question: "What is the minimum share of first-place votes a candidate needs to be taken seriously?"

- A **high** $\alpha$ (e.g., 0.1 or 0.2) acts as a strong filter. It guarantees that the winner has a non-trivial base of strong

supporters, preventing a "dark horse" candidate with no first-place votes from winning solely on high average rankings. This can increase the perceived legitimacy of the outcome.

- A **low** $\alpha$ (e.g., 0.01) makes the system more inclusive, allowing candidates with very small but dedicated followings to remain in contention. The extreme case $\alpha = 0$ removes the filter entirely.

**Guidance for $\beta$: The Consensus Safety Net**

The $\beta$ parameter acts as a crucial safety net. If the most popular Borda winner fails the $\alpha$ check, $\beta$ ensures that any alternative winner chosen by the hybrid score is still 'reasonably close' to the top consensus choice. It defines the minimum Borda score a candidate should have, relative to the maximum Borda score in the sample ($B^{\max}$), to be eligible for the hybrid calculation. An eligible candidate must have $B(p_j) \geq \beta \cdot B^{\max}$. The choice of $\beta$ is a trade-off between populism and broad appeal. It answers the question 'How far from the top consensus choice are we willing to deviate to satisfy a plurality?'

- A **high** $\beta$ (e.g., 0.9 or 0.95) strongly tethers the outcome to the Borda ranking. It allows a popular candidate to overcome a slightly more broadly acceptable one, but prevents a candidate who is deeply unpopular overall (with a very low Borda score) from winning simply by having a high plurality count. It protects against a "tyranny of the passionate minority."
- A **low** $\beta$ (e.g., 0.5) gives much more power to the $\lambda$-weighted hybrid score, allowing candidates with strong plurality support to win even if they have low overall consensus scores.

Together, these three parameters allow a community to fine-tune its collective decision-making process, creating a robust and expressive system that can be tailored to its unique social contract.

**Local Decision Logic ($\gamma, \tau_{\max}, \tau_{\min}$): Confidence and Commitment**

If the CHB parameters define the nature of a winning choice, the parameters of the `UpdateVoter` protocol, $\gamma$, $\tau_{\max}$, and $\tau_{\min}$, define the process by which a voter commits to that choice. They govern a voter's individual 'skepticism', ensuring that a `LOCK` decision is a robust commitment based on stable evidence, not a reaction to a single, potentially anomalous sample. This group of parameters allows a system designer to tune the trade-off between the speed of individual commitments and the confidence in their correctness.

The $\gamma$ parameter is the primary measure of the evidence a voter gathers before making a final commitment, assuming no early exit is triggered. Its role is to specify the total number of independent sampling rounds a voter performs to test the stability of outcomes in its local neighbourhood. The choice of $\gamma$ establishes a direct trade-off between local decision latency and local decision confidence.

- A **high** $\gamma$ (e.g., $\gamma = 20$) forces the voter to collect a large body of evidence. This increases the confidence that the resulting `LOCK` decision reflects a stable local consensus, not random noise. While this increases the message cost and time for a single voter's decision, it improves the quality of the signals sent to the rest of the network, which can accelerate global convergence by reducing dissent.
- A **low** $\gamma$ (e.g., $\gamma = 5$) allows for quicker individual decisions at a lower cost. However, these decisions are based on less evidence and are more susceptible to statistical noise, which may slow global convergence if voters commit to transient, incorrect winners.

The $\tau_{\max}$ parameter provides an optimisation for speed, allowing a voter to bypass the full $\gamma$ rounds when the local consensus is overwhelming. Its role is to define the majority threshold of round-wins required for a single candidate to trigger an immediate `LOCK`. This parameter is not independent; its value is derived from $\gamma$. The standard and recommended setting for this parameter is to define it as a simple majority of the robustness rounds: $\tau_{\max} = \lfloor \gamma/2 \rfloor + 1$. This ensures that an early exit is only possible when a candidate demonstrates clear and unambiguous dominance in the voter's samples. The trade-off for $\tau_{\max}$ is therefore linked to $\gamma$. A low $\gamma$ enables a very fast early exit (e.g., a `LOCK` after just 2 wins if $\gamma = 3$), but this decision is based on minimal evidence. Conversely, a high $\gamma$ ensures that even an "early" decision is backed by a substantial number of successful rounds.

The $\tau_{\min}$ parameter acts as a final safety check. It prevents a voter from committing to a weakly supported candidate in a highly contentious environment where no majority winner emerges after $\gamma$ rounds. Its role is to specify the minimum number of round-wins a candidate must achieve to be considered a valid plurality winner after all $\gamma$ rounds are complete. This parameter answers the question: "What is the absolute minimum level of sustained support we will accept for a winning candidate?" It establishes a trade-off between the system's lock-rate and lock-quality:

- A **high** $\tau_{\min}$ (e.g., $\tau_{\min} = \lceil \gamma/3 \rceil$) increases lock-quality by rejecting winners that lack a compelling plurality of round-wins. This reduces the risk of a voter locking on a "random" winner in a fragmented environment but increases the frequency of `NO-LOCK` outcomes, which means some voter-update cycles do not result in progress.
- A **low** $\tau_{\min}$ increases the lock-rate by being more permissive. This ensures more voter-update cycles result in a `LOCK`, but it risks introducing noise into the system if voters commit to candidates with very weak and potentially transient support, which could ultimately hinder global convergence.

**Global Termination Quorum $Q$: Defining Finality**

While parameters like $k$ and $\gamma$ govern the local decision process of a single voter, the global quorum $Q$ defines the finish line for the entire system for a given single-winner round. It answers the final question: "When can we declare that a collective decision has been reached?" Its role is to specify the fraction of the total electorate ($n$) that must `LOCK` on a single candidate before the protocol round terminates and declares that candidate the winner. An absorbing state is reached when the number of voters locked on a candidate $p_j$, denoted $N_j(t)$, satisfies $N_j(t) \geq \lceil Q \cdot n \rceil$.

The fundamental constraint, as noted in the model section, is that $Q$ must be strictly greater than $0.5$. This mathematical requirement is a critical safety guarantee, as it makes it impossible for two distinct candidates to simultaneously achieve a quorum, thus ensuring a unique winner. Beyond this constraint, the choice of $Q$ establishes a direct trade-off between the speed of finality and the strength of the result:

- A **high** $Q$ (e.g., $Q = 0.8$) sets a high bar for consensus: It requires an overwhelming supermajority of the electorate to commit to a winner, which maximises the legitimacy of the final decision. The cost is likely a longer convergence time, as the information cascade must propagate to a larger portion of the network. This setting is appropriate for high-stakes decisions where strong buy-in is essential.
- A **lower** $Q$ (e.g., $Q = 0.6$) prioritises efficiency: It allows the system to terminate as soon as a clear majority has formed, leading to faster results. This is suitable for lower-stakes decisions or in environments where speed is critical. The cost is a weaker mandate, as the decision is supported by a smaller fraction of the electorate, which might be less socially robust.

In essence, $Q$ is the parameter that codifies the community's definition of "consensus." It allows system designers to choose between demanding an undeniable supermajority or accepting a simple, efficient majority as the final word.

# F Additional Simulation Results

In this part of the technical appendix, we present additional empirical results via simulation that complement the findings in the main paper.

Figure 4 demonstrates the expressive power of the CHB rule. Using a profile designed with distinct Borda ($C_1$) and Plurality ($C_0$) winners, we show that varying $\alpha$ (popularity) and $\beta$ (consensus) creates a predictable 'policy window' where the outcome shifts between candidates, confirming CHB is tunable to a community's governance philosophy. Figure 4 (b) reveals that convergence time peaks at the decision boundaries, where the protocol must distinguish between the two nearly-equivalent viable candidates near the point of transition.

Figure 5 demonstrates the protocol robustness regarding the hybrid weight $\lambda$. Overlapping confidence intervals indicate no statistically significant difference in convergence time from pure consensus ($\lambda = 0$) to pure plurality ($\lambda = 1$). This allows communities to treat $\lambda$ as a pure governance decision without incurring a performance penalty.

Figure 6 shows the protocol's scalability with respect to the number of candidates $m$, using two complementary experimental models. Figure 6(a) tests a fixed-electorate scenario and reveals a complex, non-monotonic relationship. Initially, as $m$ increases, convergence time rises due to a contention effect, as votes are split across more viable candidates. After a peak, performance temporarily improves due to a dilution effect, where a chaotic decision space leads to faster,

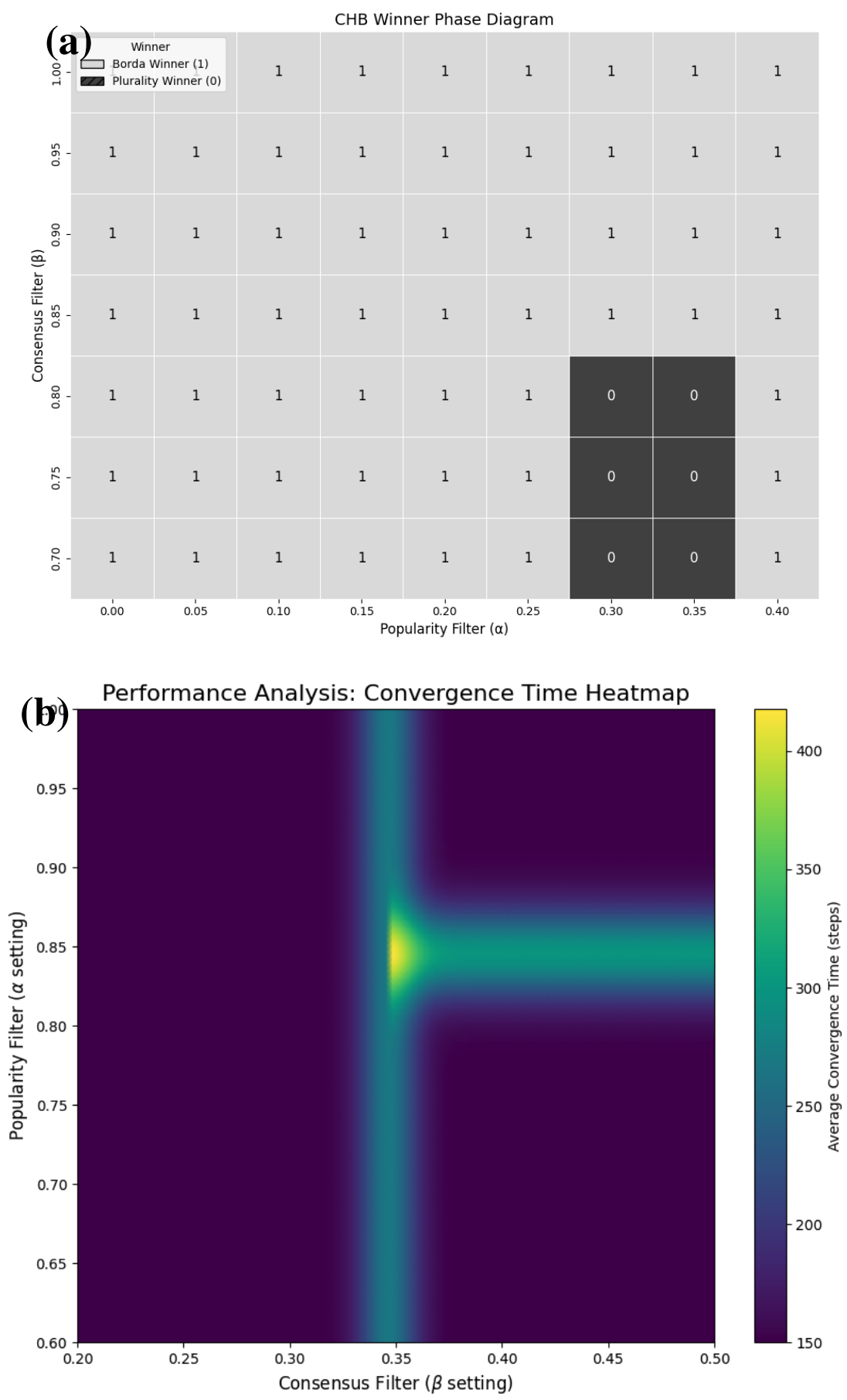

Figure 4: Impact of policy parameters $\alpha$ and $\beta$ for a profile ($n = 101$) with distinct Plurality ($t_0 = 40$, $t_1 = 35$) and Borda ($B(C_0) = 115$, $B(C_1) = 136$) winners: $(a)$ Winner outcome showing the transition across the policy window. $(b)$ Convergence time, peaking at the decision boundary, where the rule must distinguish between viable candidates.

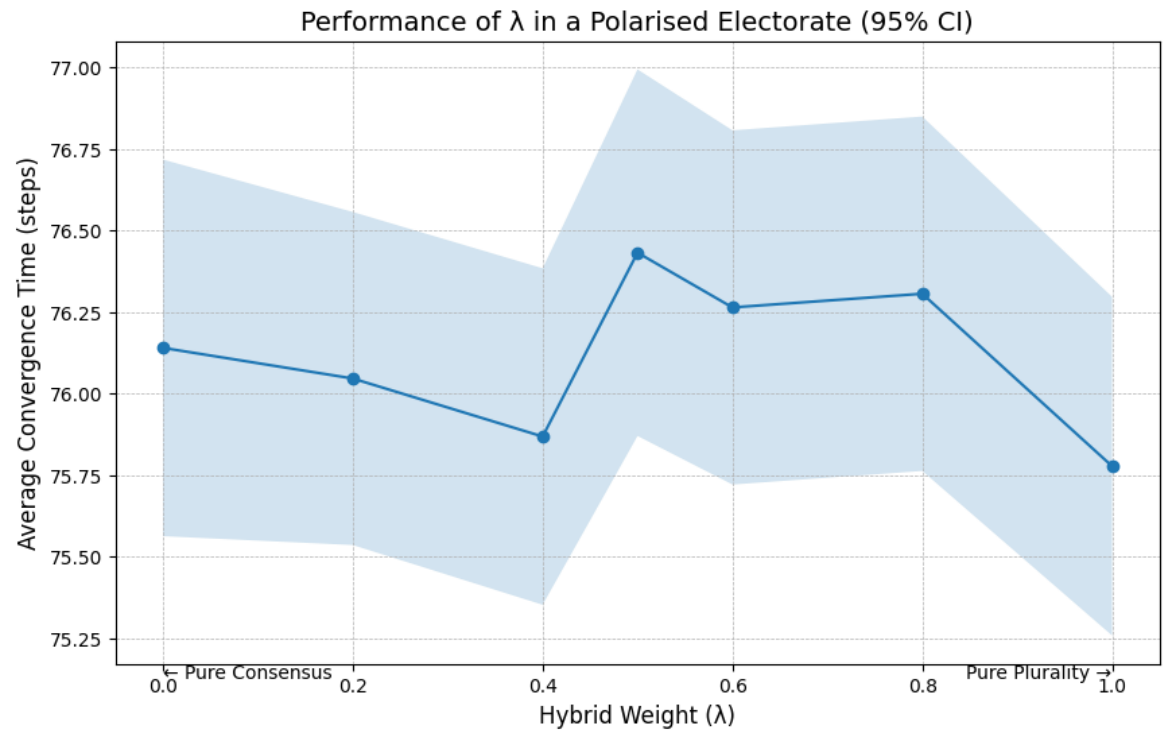

Figure 5: Convergence time vs. hybrid weight $\lambda$ in a polarised electorate (95% CI).

tie-breaker-driven outcomes. Finally, as $m$ becomes very large, the computational cost effect dominates, with the linear $O(m)$ complexity of the CHB rule causing the time to rise again.In contrast, Figure 6 (b) isolates the computational cost

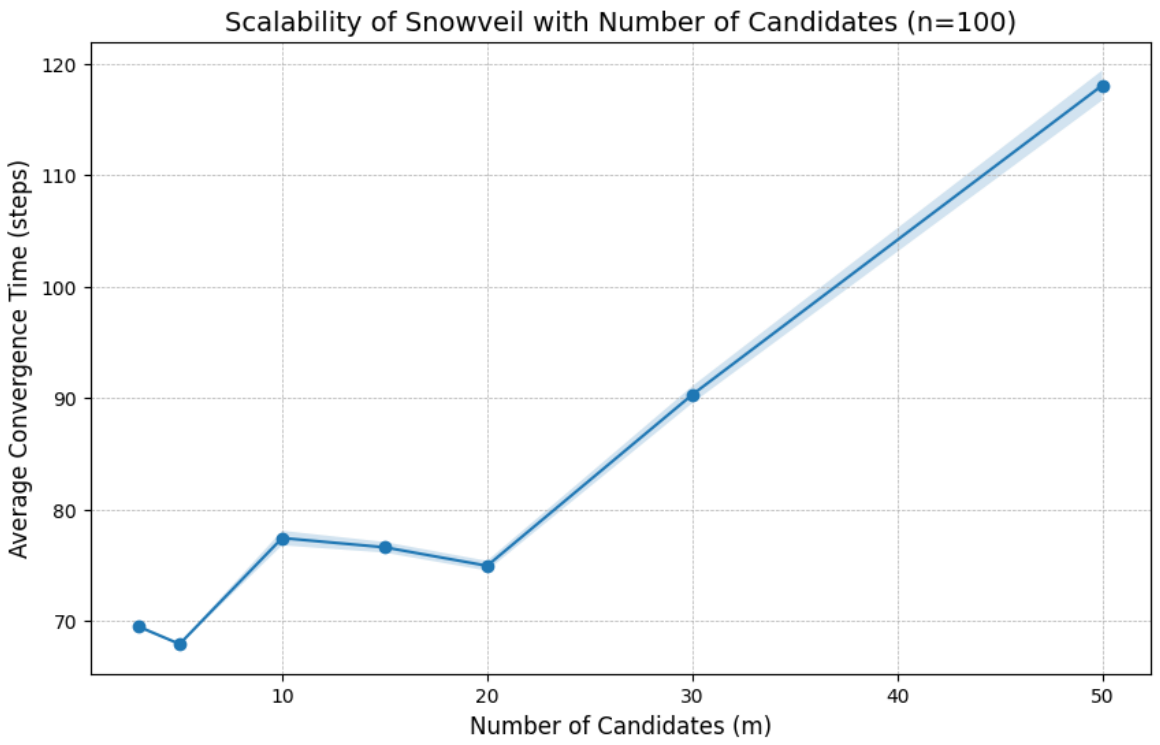

(a) Scalability with a fixed electorate size ($n = 100$).

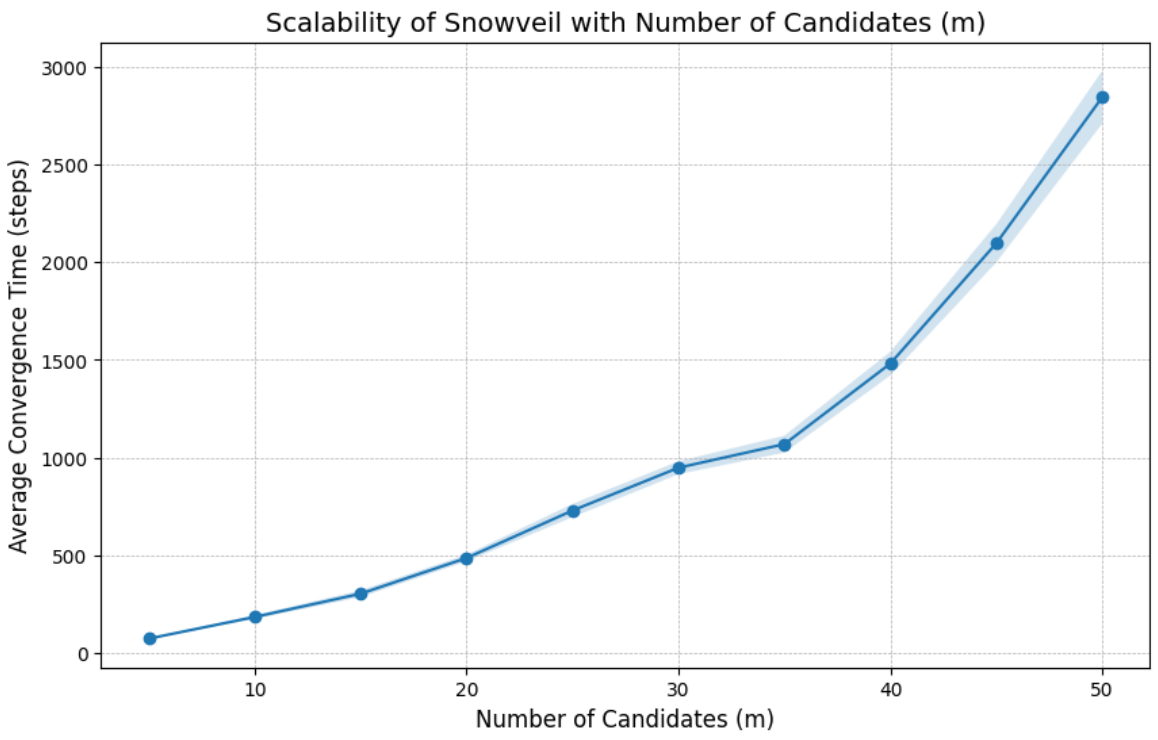

(b) Scalability with a constant voter-to-candidate ratio ($\frac{n}{m} = 20$).

Figure 6: Scalability of Snowveil with respect to the number of candidates ($m$). The two experiments reveal different dynamics of contention and computational cost.

by keeping the voter-to-candidate ratio constant. The result is a super-linear curve, driven by the $O(m)$ computational cost per step compounded by the increasing difficulty of reaching a quorum in a larger decision space. Together, the graphs highlight an interplay between social difficulty and computational complexity.